\documentclass[aps,reprint,twocolumn,superscriptaddress,nofootinbib,longbibliography]{revtex4-1}

\usepackage{amsmath, amssymb, bm, braket, dsfont, times, amsthm}
\usepackage{graphicx,xfrac}
\usepackage[dvipsnames]{xcolor}
\usepackage[breaklinks,colorlinks,allcolors=blue]{hyperref}
\usepackage{framed}
\usepackage{cleveref}
\usepackage{tikz-cd}
\usetikzlibrary{calc}
\tikzset{curve/.style={settings={#1},to path={(\tikztostart)
    .. controls ($(\tikztostart)!\pv{pos}!(\tikztotarget)!\pv{height}!270:(\tikztotarget)$)
    and ($(\tikztostart)!1-\pv{pos}!(\tikztotarget)!\pv{height}!270:(\tikztotarget)$)
    .. (\tikztotarget)\tikztonodes}},
    settings/.code={\tikzset{quiver/.cd,#1}
        \def\pv##1{\pgfkeysvalueof{/tikz/quiver/##1}}},
    quiver/.cd,pos/.initial=0.35,height/.initial=0}
\definecolor{shadecolor}{gray}{0.95}
\usepackage[normalem]{ulem}
\usepackage{mdframed}
\renewcommand{\>}{\rangle}

\newcommand{\A}{\mathcal{A}}

\newcommand{\C}{\mathcal{C}}
\newcommand{\D}{\mathcal{D}}

\renewcommand{\S}{\mathcal{S}}

\newcommand{\Z}{\mathcal{Z}}
\usepackage{enumitem}
\usepackage[mathscr]{euscript}
\newcommand\EA   {\EuScript{A}}
\newcommand\EB   {\EuScript{B}}
\newcommand\EC   {\EuScript{C}}
\newcommand\ED  {\EuScript{D}}

\newcommand\bC   {\mathbb{C}}

\newcommand\bZ  {\mathbb{Z}}



\begin{document}
\title{Classification of 1+1D gapless symmetry protected phases via topological holography}
\author{Rui Wen} \author{Andrew C. Potter} 
\affiliation{Department of Physics and Astronomy, and Stewart Blusson Quantum Matter Institute, University of British Columbia, Vancouver, BC, Canada V6T 1Z1}
\date{\today}                                       

\begin{abstract}
Symmetry topological field theory (SymTFT) gives a holographic correspondence between systems with a global symmetry and a higher-dimensional topological field theory. In this framework, classification of gapped phases of matter in spacetime dimension 1+1D correspond to classifications of mechanisms to confine the SymTFT by condensing anyons. 
In this work, we extend these results to characterize gapless symmetry-protected topological states: symmetry-enriched gapless phases or critical points that exhibit edge modes protected by symmetry and topology. 
We establish a one-to-one correspondence between 1+1D bosonic gSPTs, and partially-confined boundaries of 2+1D SymTFTs. From general physical considerations, we determine the set of data and consistency conditions required to define a 1+1D gSPT, and show that this data precisely matches that of symmetry-preserving partial confinement (or partially gapped boundaries) of 2+1D quantum double models. 
We illustrate this correspondence through a dimensional reduction (thin-slab) construction, which enables a physically-intuitive derivation of how properties of the gSPT such as edge modes, emergent anomalies, and stability to perturbations arise from the SymTFT perspective.
\end{abstract}
\maketitle

\tableofcontents
\vspace{24pt}
Symmetry plays a crucial role in modern condensed matter physiscs. In the presence of symmetry short range entangled states can be enriched and split into inequivalent quantum phases, known as symmetry protected topological (SPT) phases~\cite{Chen_2011_1,Chen_2011_2,Chen_2011_3,Gu_2009,Pollmann_2010,Turner_2011,Fidkowski_2011,Else_2014,Chen_2013,Wang_2014,Kapustin_2015,Senthil_2015}. SPT states have short-range entangled gapped bulks, but exhibit anomalous edge states that are confined to the boundary by the bulk gap.
Recent works~\cite{Kong_2020,Chatterjee_2023_1,Chatterjee_2023_2,moradi2022topological,freed2023topological,Fuchs_2002,Apruzzi_2023,kaidi2022symmetry,freed2023introduction,kaidi2023symmetry,zhang2023anomalies,bhardwaj2023gapped,Lichtman_2021,Apruzzi_2022,Lin_2023,cordova2023anomalies} have proposed a topological holographic principle between gapped 1+1D systems with symmetry group $\Gamma$, and anyon condensations of a symmetry topological field theory (SymTFT): a 2+1D topological order corresponding to a quantum double $D(\Gamma)$. 
This works conjectured that may be possible to utilize the dual topological order to fully classify and characterize 1+1D gapped as well as gapless states in a unified way. 
While the classification of low-dimensional gapped phases is well understood by other methods, situation for gapless phases and critical points are much less well understood, and any sharp, non-perturbative, topological insights into gapless phases would be valuable to theory. Moreover, SymTFT has been shown to be powerful in analyzing symmetries of gapped systems, including for systems with generalized, non-invertible symmetries.

Despite the importance of the bulk gap in defining conventional topological phases, it has also been recently shown that topological properties such as non-local order parameters, localized edge modes, and emergent anomalies can also arise in \emph{gapless} systems
~\cite{Keselman_2015, Scaffidi_2017,Verresen_2018, Verresen_2021,Thorngren_2021,li2023decorated,li2023intrinsicallypurely,Wen_2023,Parker_2018,mondal2023symmetryenriched,Hidaka_2022,Ma_2022,Borla_2021,Yang_2023,Zhang_2023}.
For gapless phases or critical points in which these topological features rely on symmetry protection have been dubbed gapless SPT (gSPT) states.
Multiple families of gSPTs have been identified in the literature.
Weak gSPTs arise as critical points between SPT and spontaneous symmetry breaking (SSB) phases, and exhibit a partial set of edge modes from the adjacent gapped SPT~\cite{Verresen_2021,Senthil_2015,li2023decorated}. In particular, weak gSPT edge modes can be removed by stacking the system with a gapped SPT.
Other gSPTs are \emph{intrinsically gapless} (igSPTs), and have topological features that would be forbidden in any gapped state~\cite{Thorngren_2021,li2023intrinsicallypurely,Wen_2023}, such as emergent bulk symmetry anomalies that protect edge modes which could not arise at the boundary of a gapped SPT with the same symmetry.

Given the success of SymTFT in characterizing gapped systems, it is natural to ask if one could utilize the SymTFT to characterize gSPTs, to provide new insights into the topological properties of gSPTs. To achieve this goal it is crucial to understand how the topological features of gSPTs such as edge modes and emergent anomalies enter the paradigm of SymTFT. 
 In this work we address these questions and develop a framework for describing gSPTs via SymTFT. To formally establish this framework, we provide a theory of 1+1D bosonic gSPTs that includes a set of data determing the algebra of symmetry operators and the consistency conditions these data must satisfy. The data consists of two phase factors $(\eta,\epsilon)$, where $\eta$ specifies the SPT class of the gapped sector of the gSPT and $\epsilon$ determines the interplay between the gapped sector and the gapless sector. We show that these data fully characterizes the topological features of a gSPT including its edge modes and emergent anomaly. Analyzing the structure of the emergent anomaly, we also re-derive a bulk-edge (1+1D/0+1D) correspondence for gSPTs~\cite{Thorngren_2021,Wen_2023}: the edge modes determine the bulk emergent anomaly and the bulk emergent anomaly determines the edge modes up to stacking with a gapped SPT. Notably, precisely the same set of data is required to specify a symmetric, partially-confining condensation of a quantum double. Moreover, an anyon condensation in a quantum double typically results in a twisted quantum double, and we find that the resulting post-condensation quantum double's twist matches the emergent anomaly of the corresponding gSPT. These findings lay the foundation of incorporating gSPTs into the SymTFT paradigm. 
We further directly illustrate the SymTFT description of gSPTs via a dimensional reduction procedure in which we consider a thin slab of the SymTFT with gapped boundaries specified by some anyon condensation. This construction provides a physically intuitive, and graphical method to constrain the phase diagram of the dual 1+1D system. We use this technique to analyze the stability of gSPTs and their relation to other phases in the phase diagram.

This work is organized as follows. In Section~\ref{sec:bg} we review basics of gSPTs and the SymTFT concept. 
Along the way, we generalize a thin-slab dimensional reduction procedure~\cite{Thorngren_2021} to describe interfaces between gapped SPT and trivial symmetric states, which enables us to reproduce the SPT edge states and their projective symmetry action via the braiding properties of anyon line operators in the SymTFT dual.
 In Section~\ref{sec:z4igspt_symtft} we initiate the attempt to describe gSPTs by SymTFT. We use the $\bZ_4$-igSPT as an example and provide a SymTFT construction that is able characterize properties of the $\bZ_4$-igSPT. We then develop the general theory of SymTFT/gSPT correspondence. We first analysis the structure of 1+1D gSPTs in Section~\ref{sec:gspt_structure} and provide a classification. Then in Section~\ref{sec:general_condense} we show that the symmetric, partially-confining condensations in 2+1D quantum doubles have exactly the same structure as that of 1+1D gSPTs. We utilize the established duality to study two families of gSPTs and their SymTFT description in Section ~\ref{sec:gSPTegs}. Finally we conclude with a discussion of possible routes to generalize the SymTFT/gSPT correspondence to higher dimensions and generalized higher-form or non-invertible symmetries.

\section{Background: gapless SPTs, and SymTFT\label{sec:bg}}

\subsection{Gapless SPTs: Generalities}

Gapless SPTs are examples of SECs, in which one cannot locally distinguish between a ``trivial" critical point and a symmetry-enriched one from local bulk measurements. Rather, the differences between different gapless SPTs are only evident in non-local bulk probes, or local edge probes.
One can make this notion more specific by analogy to gapped SPTs. Different gapped $G$-SPT ground-states can be connected by a finite-depth local unitary (FDLU), $U$, that is overall symmetric ($[U,g]=0~\forall g\in G$) but which is not symmetrically generated ($U\neq e^{-iH}$ for any local $G$-symmetric $H$). This definition cannot be directly ported to the gapless setting as even different instances of a gapless state with the same universal scaling properties, e.g. two instances of a conformal field theory (CFT) perturbed by different irrelevant operators, cannot be connected by an FDLU. However, we can generalize the notion of symmetric FDLU-(in)equivalence by defining i) that two ground-states are in the same universality class if they flow to the same RG fixed point after applying an overall symmetric FDLU, ii) ground-states with the same universality class are distinct gapless SPT classes if they cannot be connected in this way by any symmetrically-generated FDLU.
An immediate corollary of this definition is that local scaling operators in distinct gapless SPTs of the same type of criticality have the same symmetry properties as conjugating a local operator with definite symmetry quantum number with an overall-symmetric $U$ preserves its symmetry quantum number. However, the symmetry properties of non-local scaling operators, such as a disorder operator that inserts a domain wall, may change under such an overall symmetric FDLU, leading to distinct classes of gapless SPTs.  These notions can be made more precise for 1+1D conformal field theories (CFTs)~\cite{Verresen_2021}, for which the data specifying a universality class is well understood and characterized by the spectrum and fusion rules for primary scaling operators.

\subsubsection{Example of a 1+1D igSPT:}
A simple igSPT with full symmetry $\Gamma=\bZ_4$ was constructed in studied in~\cite{li2023decorated,Wen_2023}. We will use this example extensively throughout this paper to illustrate the formal results.
A lattice Hamiltonian realizing the $\bZ_4$-igSPT chain is given by~\cite{li2023decorated}
\begin{align}
    H_{\bZ_4-\text{igSPT}}&=H_0+H_\Delta 
    \nonumber\\
    H_0&=-g\sum_i\left(\tau^z_{i-1/2}\sigma^x_i\tau^z_{i+1/2}+\tau^y_{i-1/2}\sigma^x_i\tau^y_{i+1/2}\right)
    \nonumber\\
    H_\Delta &= \Delta \sum_i \sigma^z_{i-1}\tau^x_{i-1/2}\sigma^z_i
\end{align} 
with two species of spins per unit cell represented by Pauli operators $\sigma^\alpha, \tau^\alpha$. The system has a $\bZ_4$ symmetry generated by 
\begin{align}
    U_s=\prod_i \sigma^x_ie^{\frac{i\pi}{4}(1-\tau ^x_{i+1/2})}
\end{align}
Below the energy scale, $\Delta$ (taken to be very large), $\tau^x_{i-1/2}$ is locked to $\sigma^z_{i-1}\sigma^z_{i}$, and the symmetry effectively reduces to an \emph{anomalous} $G=\Z_2$ symmetry:
\begin{align}
    U_s\approx \prod_i \sigma^x_ie^{\frac{i\pi}{4}(1-\sigma^z_{i}\sigma^z_{i+1})}
\end{align}
Viewed as a $\bZ_2$ symmetry, this symmetry action is anomalous, and has the same anomaly as the edge of a 2+1D SPT phase (the Levin-Gu phase~\cite{Levin_2012}). This emergent anomaly prevents the system from realizing a symmetry-preserving short-range entangled state. Instead, the only options are to spontaneously break the symmetry, or form a gapless state. The above choice of $H_0$ realizes the latter option, reducing at low energies to the critical spin-chain:
\begin{align}
H_0\approx-\sum_i \sigma^x_i-\sigma^z_{i-1}\sigma^x_i\sigma^z_{i+1}.
\end{align}

Further, in an open chain, one finds that acting twice with the low energy symmetry operator ``pumps" a $G=\Z_2$ charge of the low energy symmetry group onto each end of the boundary: $U_s^2 = \sigma^z_1\sigma^z_L$, where $1,L$ are the first and last sites in the chain respectively~\cite{li2023decorated,Wen_2023}. This symmetry-charge pumping, locally anticommutes with the $U_s$ symmetry, protecting a two-fold ground-state degeneracy, split only exponentially-weakly in system size, $L$ by finite size effects (in contrast to polynomial scaling with $1/L$ for gap to bulk excitations). Away from this idealized fixed point, this ground-state degeneracy is robust, so long as symmetry remains intact, and the gap to the $\tau^z$ particles remains open.

Formally, the structure of the emergent anomaly is captured by the short-exact sequence of groups:
\begin{align}
    1\rightarrow A\xrightarrow[]{j} \Gamma\xrightarrow[]{p} G\rightarrow 1
\end{align}
which defines a group extension of $G$ by $A$ to $\Gamma$. 
Microscopically (in the UV), the system has full symmetry $\Gamma$, but excitations carrying charges of the normal subgroup $A\trianglelefteq \Gamma$ are gapped out by $H_\Delta$. Therefore, below the energy scale of this gap, the symmetry is effectively reduced to the quotient $G=\Gamma/A$. We call $A,G$ the gapped symmetry and the gapless symmetry respectively. The gapless symmetry $G$ can be anomalous and the anomaly is represented by a cocycle $\omega\in H^{d+2}(G,U(1))$. 
In the above example, $A,G=\mathbb{Z}_2$, and $\Gamma=\bZ_4$, and $\omega(a,b,c) = (-1)^{abc}$ is the anomaly of the edge of a 2+1D Levin-Gu phase~\cite{Levin_2012}.
Since the full symmetry $\Gamma$ is not anomalous, the emergent anomaly must be ``lifted" by the gapped sector. Formally, this implies that ,upon pull-back by the projection $p$, the cocycle becomes a coboundary: $p^*(\omega)=d\alpha$. Moreover, the system can be put in a $\Gamma$-symmetric gapped phase. However, the anomaly in the IR has the consequence that one must first close the gap to the $A$-degrees of freedom before tuning the system to a gapped symmetric state. Thus the nontrivial topological properties of the system are protected by the $A$-gap, despite the presence of gapless $G$-charged excitations. 

\subsection{The Symmetry topological field theory (SymTFT) concept \label{sec:symtft}}
SymTFT is a method of studying a given symmetry via a higher dimensional topological order. SymTFT has been proven powerful in classifying SPTs protected by generalized symmetry, identifying anomalies of non-invertible symmetries, revealing dualities between different phases, and beyond~\cite{moradi2022topological,freed2023topological,Fuchs_2002,Apruzzi_2023,kaidi2022symmetry,freed2023introduction,kaidi2023symmetry,zhang2023anomalies,bhardwaj2023gapped,Lichtman_2021,Apruzzi_2022,Lin_2023,cordova2023anomalies}.  We here review the construction of SymTFT for ordinary internal symmetries (invertible, 0-form symmetries).

For a system with a symmetry group $G$, the SymTFT is a $G$-gauge theory in one higher dimension. This correspondence can be motivated in two different ways: 
First, there is a relation between global gauge transformations, and a boundary global symmetry, as we now review. Though the SymTFT correspondence has only been established for discrete groups $G$, consider for a moment the more familiar $U(1)$ gauge theory on a space $\Sigma$, with electric field $E$, and charge density $\rho$. By Gauss' law, $\nabla\cdot E = \rho$, a global gauge transformation, reduces to a $U(1)$ symmetry transformation that acts only on the boundary, $\partial \Sigma$ of the system: 
$U_\theta = e^{i\theta \int_\Sigma \rho} = e^{i\theta \int \nabla\cdot E} = e^{i\int_{\partial\Sigma} E}$.
This boundary symmetry of a gauge theory is also called asymptotic symmetry~\cite{harlow2019symmetries, PhysRev.128.2851,Strominger_2014,strominger2018lectures} and plays a crucial role in the recently proposed ``Higgs=SPT'' correspondence~\cite{thorngren2023higgs,verresen2022higgs}. This relation between bulk gauge symmetry and edge global symmetry also hold for discrete symmetries.
Thus, a $G$-symmetric system can always be embedded into the boundary of a $G$ gauge theory in one higher dimension. We note that, if the $G$ symmetry action is anomalous, the corresponding bulk gauge theory would be a twisted gauge theory, also known as the Dijkgraaf-Witten theory with a nontrivial cocycle.

A second complementary perspective, is that one can recover a $G$-symmetric system in $d+1$D from a dimensional reduction procedure by considering a thin slab of $(d+1)+1$D (twisted) $G$ gauge theory, with appropriate open boundary conditions in the thin direction, and periodic boundary conditions in the other directions. 
We illustrate this dimensional reduction for the simplest case of a 2+1D $\Z_2$ gauge theory (toric code topological order) with anyon types $e,m,f=e\times m$. Here, consider the top boundary of the slab to be a fixed reference gapped state. Gapped boundaries of topological orders correspond to anyon condensations that fully confine the topological order~\cite{kong2014anyon,Kitaev_2012}. Here, we consider the reference top boundary state as an $e$-condensed boundary. 
Define a global $\bZ_2$ symmetry operation as nucleating a pair of $m$ particles, and dragging them around the periodic cycle. Gapped states of this quasi-1d system correspond to different gapped boundaries for the bottom of the slab. The $m$ condensed boundary is symmetric under the above-defined symmetry, since it can absorb the $m$ line that generates the symmetry. By contrast, the $e$ condensed bottom boundary corresponds to a spontaneous symmetry broken state, which has a dual symmetry under pulling an $e$ particle from the top-boundary condensate and moving it to the bottom boundary condensate. This dual symmetry anti-commutes with the original $\bZ_2$ symmetry, due to the mutual semionic statistics of the $e$ and $m$ particles, leading to a two-fold degenerate ground-space. 
This structure of the global $\bZ_2$ m-symmetry and dual $\bZ_2$ e-symmetry, matches that of the lattice spin-1/2 Ising model with global $\bZ_2$ symmetry $g_m=\prod_i \sigma^x_i$ respected by symmetric (paramagnetic) short range entangled phases, and dual domain wall conservation symmetry: $g_e=\prod_i \sigma^z_i\sigma^z_{i+1}$ respected in symmetry broken phases: these two symmetries anticommute when restricted to overlapping finite intervals, and were referred to as ``categorical symmetries" or symmetry topological order (SymTO) in~\cite{Kong_2020,Chatterjee_2023_1,Chatterjee_2023_2,inamura202321d,chatterjee2023emergent}.
The Ising critical point separating the two phases also corresponds to the phase transition between the $e$ and $m$ condensed bottom boundary~\cite{Lichtman_2021,Chatterjee_2023_2}.

In 1+1D systems, this structure generalizes to any finite internal unitary symmetries. In the broken phase of such a symmetry, a dual symmetry appears that encodes the conservation of domain wall excitations. The original symmetry and the dual symmetry together form the categorical symmetry of the system and can be uniformly described by a 2+1D topological order. As we have seen in the toric code example, the symmetry and the dual symmetry correspond to anyon string operators of gauge flux and gauge charge respectively. The statistics of anyons in the 2+1D topological order then encodes the commutation relations between the symmetry and the dual symmetry. The thin-slab construction is a way to make the categorical symmetry manifest: if the 1+1D system has a symmetry $G$ with anomaly $\omega\in H^3[G,U(1)]$, then one can always put the system on the boundary of the twisted gauge theory (a.k.a twisted quantum double) $D_\omega(G)$. The original/dual symmetry is then realized by dragging fluxes/charges around the periodic direction. To specify the symmetry and the dual symmetry, one fixes the top boundary to be one that condenses all the charges. Different phases of the 1+1D system are therefore realized by choosing different boundary conditions on the bottom boundary. 

This thin-slab construction, which can be viewed as a concrete realization of the SymTFT idea, translates the question of classifying phases of matter with symmetry in 1+1D into the question of classifying boundaries of 2+1D topological orders. The latter question is well-understood: Boundaries of 2+1D topological orders are completely determined by anyons that condense at the boundary\cite{kong2014anyon,Kong_2017,Kong_2020_2,Kong_2021,Kitaev_2012}. For abelian topological orders, an anyon condensation is specified by a group of self and mutual bosons $\A$. After condensing $\A$ the anyons that have non-trivial braiding with $\A$ become confined and the topological order is reduced to one whose excitations only include the deconfined anyons that braid trivially with $\A$. The theory of anyon condensation for generic topological orders is more involved and we defer the discussion to section~\ref{sec:general_condense}.
\begin{figure*}
    \centering
    \includegraphics[width=1\textwidth]{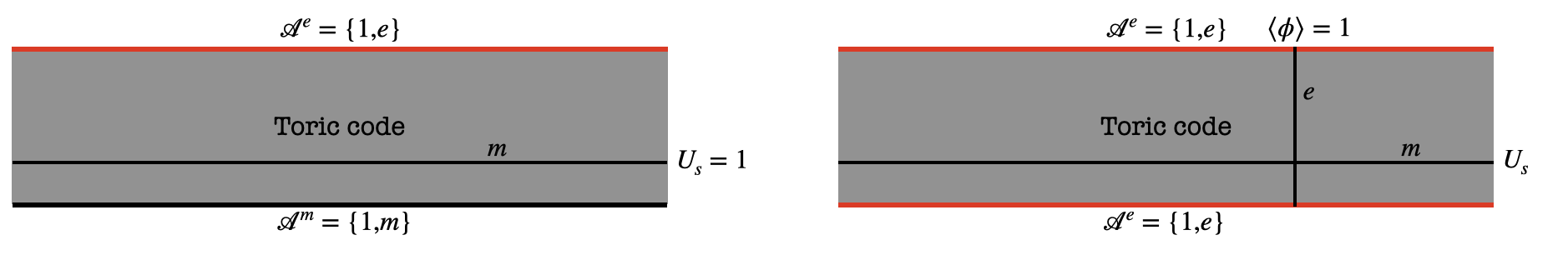}
        \caption{\textbf{SymTFT description of $\bZ_2$ symmetry.} The thin slab hosts toric code in its bulk. On the top reference boundary we fix the $e$-condensed boundary condition. The horizontal $m$-line becomes a global symmetry for the effective 1+1D system which we identify as the generator of a $\bZ_2$ symmetry. In the left figure we consider condensing $m$ on the physical boundary. In this case the horizontal $m$-string can be absorbed by the physical boundary and acts as identify. The effective 1+1D system is therefore gapped and symmetric. In the right figure we consider condensing $e$ on the physical boundary. In this case the vertical $e$-string acts as the order paramater of the symmetry and has non-zero expectation value, signaling SSB. Equivalently, the system is now invariant under the dual symmetry represented by a horizontal $e$-string.}
        \label{fig:z2 example}
\end{figure*}
\begin{figure*}
    \centering
    \includegraphics[width=1\textwidth]{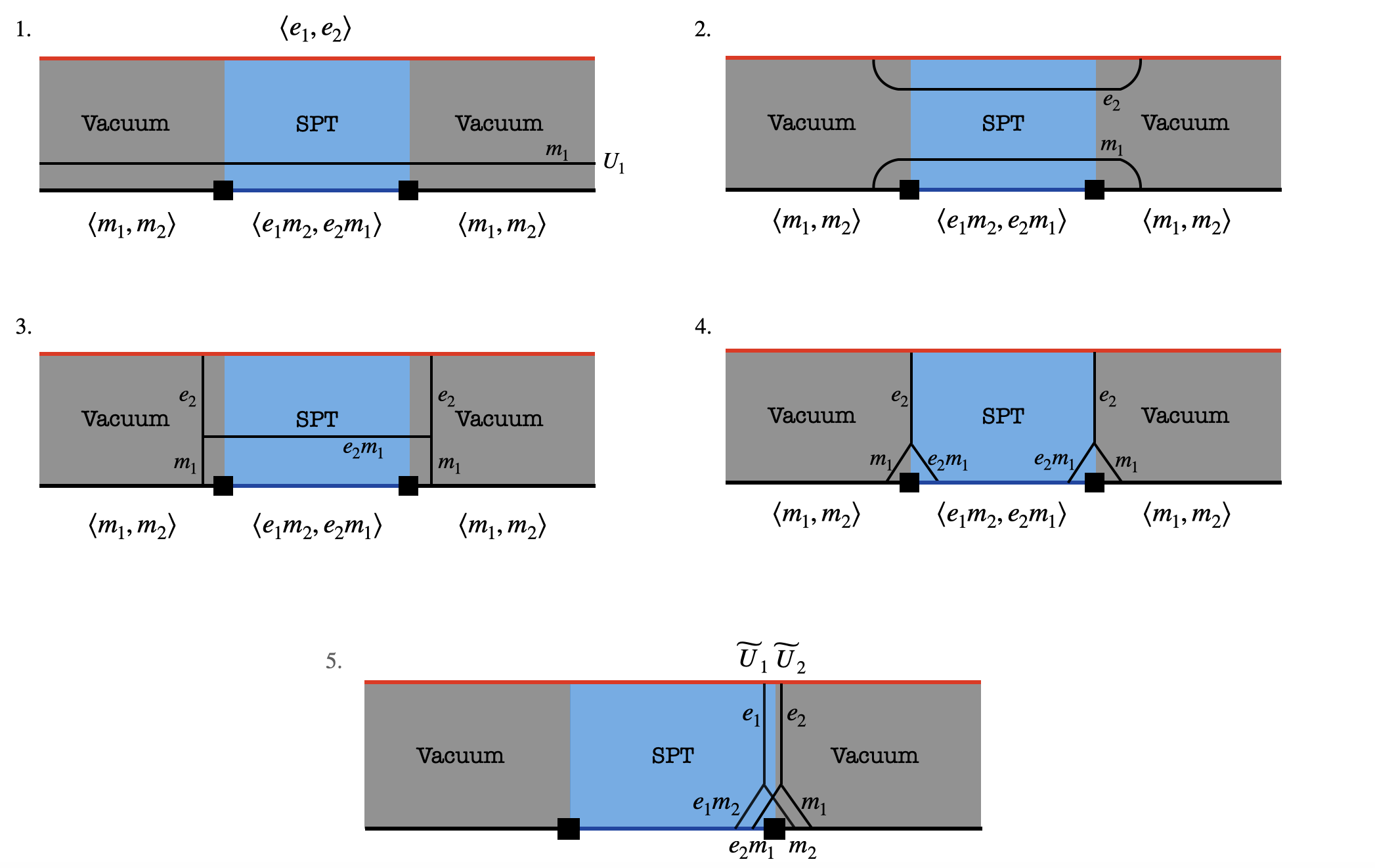}
        \caption{\textbf{SymTFT description of the cluster chain.} 1.The condensation $\{e_1m_2,e_2m_1\}$ describes the SPT state protected by $\bZ_2^A\times \bZ_2^B$. The $m_1$-line operator is the generator of $\bZ_2^A$. 2. we pull out an $e_2$-string from the $e_2$-condensation at the top boundary, which acts as identity on the ground state. 3.We multiply the $e_2$-string and the $m_1$ string to obtain an $H$-shape operator. 4. The middle part of the $H$-shape operator is absorbed by the bottom $m_1e_2$-condensation, leaving an operator that is supported at the left and right edge of the SPT. Similar deformation can be perfomed for $\bZ_2^B$ action. Therefore the symmetry actions of the SPT can be localized to the edge in the ground space. 5. The localization of $\bZ_2^A$ action and $\bZ_2^B$ action to the right edge of the SPT is shown. Since the two operators intersect at one point, the mutual $\pi$ statistics between $e_2m_1$ and $m_1$ indicates that the two operators anti-commute. Therefore the edge symmetry actions form a projective representation of $\bZ^A_2\times \bZ_2^B$.
        }
        \label{fig:SPT_edge1}
\end{figure*}
In general we can devide anyon condensations into two families: If the condensation completely confines the theory and no anyons other than vacuum stay deconfined, then we call the condensation fully-confining, otherwise we call the condensation partially-confining. In the boundary/anyon condensation dictionary of 2+1D topological orders~\cite{Kong_2017,Kong_2020_2,Kong_2021}, gapped boundaries corresond to full-confining condensations and gapless boundaries corresond to partially-confining condensations. It is then clear that in the thin-slab construction, gapped/gapless phases of the 1+1D system is obtained by choosing fully/partially confining condensations on the bottom boundary. Moreover, if on the bottom boundary any of the gauge charges is condensed, then the effective 1+1D system is invariant under a dual symmetry implying the original symmetry is at least partially broken. Therefore in order to describe a fully symmetric 1+1D system the bottom boundary must condense no charges. Motivated by this relation between charge condensation and symmetry breaking, we call a condensation symmetric if it does not condense any gauge charges. 

The SymTFT is known to be capable of characterizing gapped phases. We have seen in the $\bZ_2$ symmetry example that the dual toric code has three boundary conditions corresonding to exactly three different phases of the $\bZ_2$ symmetry. In 1+1D there are also nontrivial SPT phases that have edge modes protected by symmetry. The SymTFT is also capable of describing edge modes of SPTs, as we discuss now. To illustrate the principle we consider the symmetry $\bZ_2\times\bZ_2$ that supports a non-trivial SPT known as the cluster chain. The SymTFT in this case is the quantum double of $\bZ_2\times \bZ_2$, which can be viewed as two copies of toric code. We label the anyons in each copy by $e_i,m_i,f_i,i=1,2$. Since the cluster chain is a gapped symmetric system, it should be dual to a symmetric, fully-confining condensation of $D(\bZ_2\times \bZ_2)$. Consider the option $\A^{SPT}=\langle e_1m_2,e_2m_1\rangle$ that satisfies all the conditions. We then consider a thin-slab construction where on the top boundary we condense all charges $e_1,e_2$, and on the bottom boundary there is an interface between $\A^{SPT}$ condensation and $\A^{m}=\langle m_1,m_2\rangle$ condensation. The set up is illustrated in Fig.~\ref{fig:SPT_edge1}. As shown in Fig~\ref{fig:SPT_edge1}, the symmetry of the system, in this case a horizontal $m_1$ or $m_2$ string, can be localized to the edge of the SPT and the localized symmetry actions form a projective representation of $\bZ_2\times \bZ_2$. Therefore the condensation $\A^{SPT}$ corresponds to a gapped $\bZ_2\times\bZ_2 $ symmetric 1+1 system with non-trivial edge modes, i.e. the cluster state.

\section{SymTFT for GSPTS?\label{sec:z4igspt_symtft}}
SymTFT has been applied widely to analyze gapped phases. As we have seen, it is able to characterize both symmetric and symmetry-breaking phases. In the symmetric case, SymTFT has been shown known to be able to reproduce the classification of SPTs~\cite{zhang2023anomalies}. Given this capacity, one might wonder if SymTFT could shed light on gapless phases. A preliminary exploration of this idea was performed in~\cite{Chatterjee_2023_2}, where the authors conjectured that SymTFT is capable of classifying 1+1D gapless phases with symmetry. To elucidate, the thin slab construction indicates that a symmetric gapless 1+1D system corresponds to a symmetric, partially-confining condensation on the bottom boundary. Thus, it's plausible that all symmetric, partially-confining condensations of a 2+1D quantum double $D(G)$ represent the classification for all possible 1+1D $G$-symmetric gapless systems. Furthermore, a local low-energy equivalence principle (LLEP) was introduced in~\cite{Chatterjee_2023_2}. If the topological order obtained by condensing $\A$ in the topological order $D(G)$ is represented as $D(G)/\A$, this principle suggests that the local low-energy properties of a 1+1D system dual to a condensation $\A$ of $D(G)$, are equivalent to those of a 1+1D system dual to the trivial condensation $1$ of $D(G)/\A$. Using  LLEP, we can derive the local low-energy properties of 1+1D systems dual to a given partially-confining condensation of the SymTFT. As an example, by condensing the double flux $m^2$ in $D(\bZ_4)$, we obtain a toric code topological order $D(\bZ_2)$ formed by $e^2$ and $m$. Thus, the condensation $m^2$ corresponds to a 1+1D system exhibiting local low-energy properties identical to the Ising critical point.

However, the LLEP does not fully clarify the correspondence between 1+1D phases and anyon condensation of 2+1D quantum doubles. For example, there could be multiple condensations giving rise to equivalent post-condensation topological order, then LLEP suggests that these condensations describe 1+1D systems with identical local low energy properties. However there are non-local properties of 1+1D systems that can be used to distinguish different phases, the LLEP does not clarify how these non-local properties are described in the anyon condensation picture. As a more concrete example, all SPT phases associated with a given symmetry have the same local low energy properties: they are adiabatically connected to a trivial product state and there is no local order parameter that can distinguish one SPT from another. However different SPT phases are inequivalent quantum phases if symmetry is to be preserved and they can distinguished by non-local string order parameters. Therefore the LLEP must only be part of a more complete dictionary. In the gapped phases/fully-confining condensation case this is well-understood: different symmetric full-confining condensations correspond to different 1+1D SPT phases. However it is still unclear what is the complete 1+1D correspondence for a partially-confining condensation other than what the LLEP provides.

On other hand, gSPTs are gapless systems with symmetry, therefore one might wonder how does gSPT fit into the proposed SymTFT description of gapless systems? Specifically, are there symmetric partially-confining condensations that are dual to gSPTs? And if there are, how does anyon condensation describe the properties of a gSPT, such as the symmetry extension structure, emergent anomaly, and edge modes? We address these questions in the rest of this paper and establish a complete framework for describing gSPTs by SymTFT. We will see that symmetric partially-confining condensations are exactly dual to gSPTs, thereby solidify the SymTFT for gapless systems conjecture. To illustrate, let us examine the $\bZ_4$-igSPT as a case study. 
\subsection{SymTFT for the $\bZ_4$-igSPT}

\begin{figure*}
    \centering
    \includegraphics[width=1.2\columnwidth]{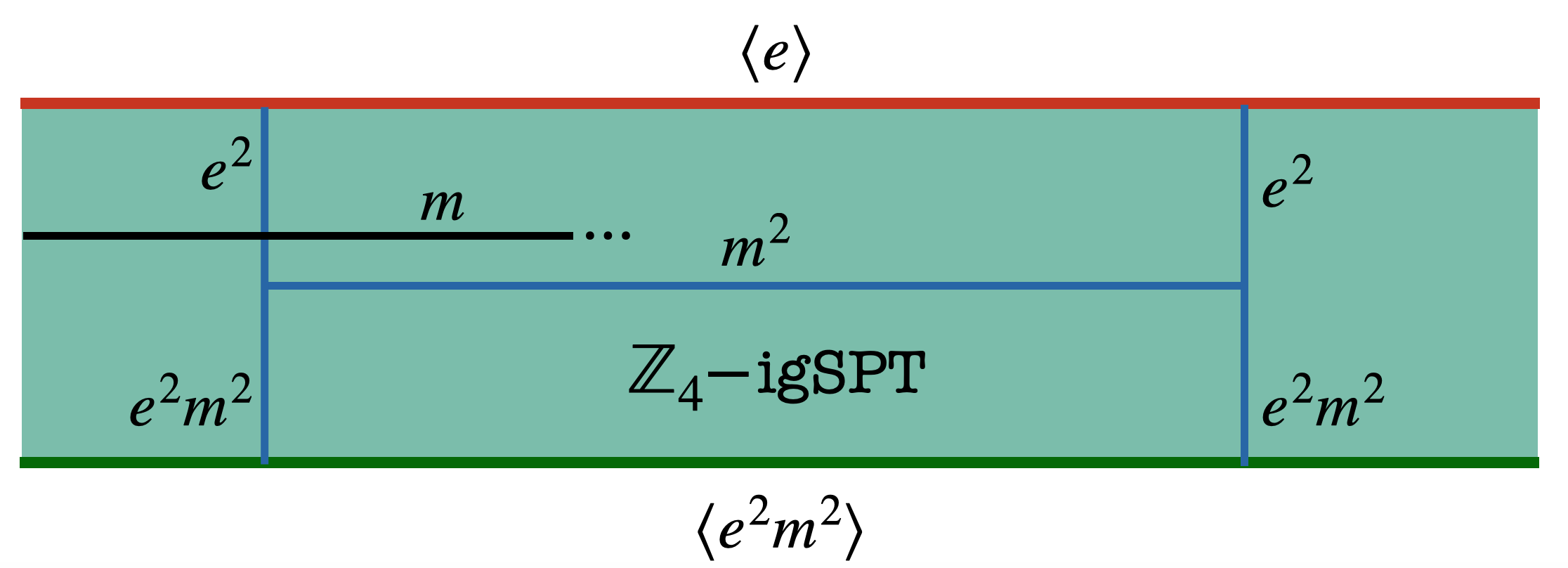}
    \caption{\textbf{String order parameter of the $\bZ_4$-igSPT in the thin slab construction.}  The slab describes an effective 1+1D system that we identify as the $\bZ_4$-igSPT. On the top boundary we condense the charge $e$, on the bottom boundary we condense $e^2m^2$. The $H$-shaped operator in blue represents the string order parameter of the $\bZ_4$-igSPT. It is a decorated $U_{s^2}$ action, with ends charged under $U_s$, i.e. a horizontal $m$-string.}
    \label{fig:igSPT_string}
    \end{figure*}
\begin{figure*}
    \centering
    \includegraphics[width=1.2\columnwidth]{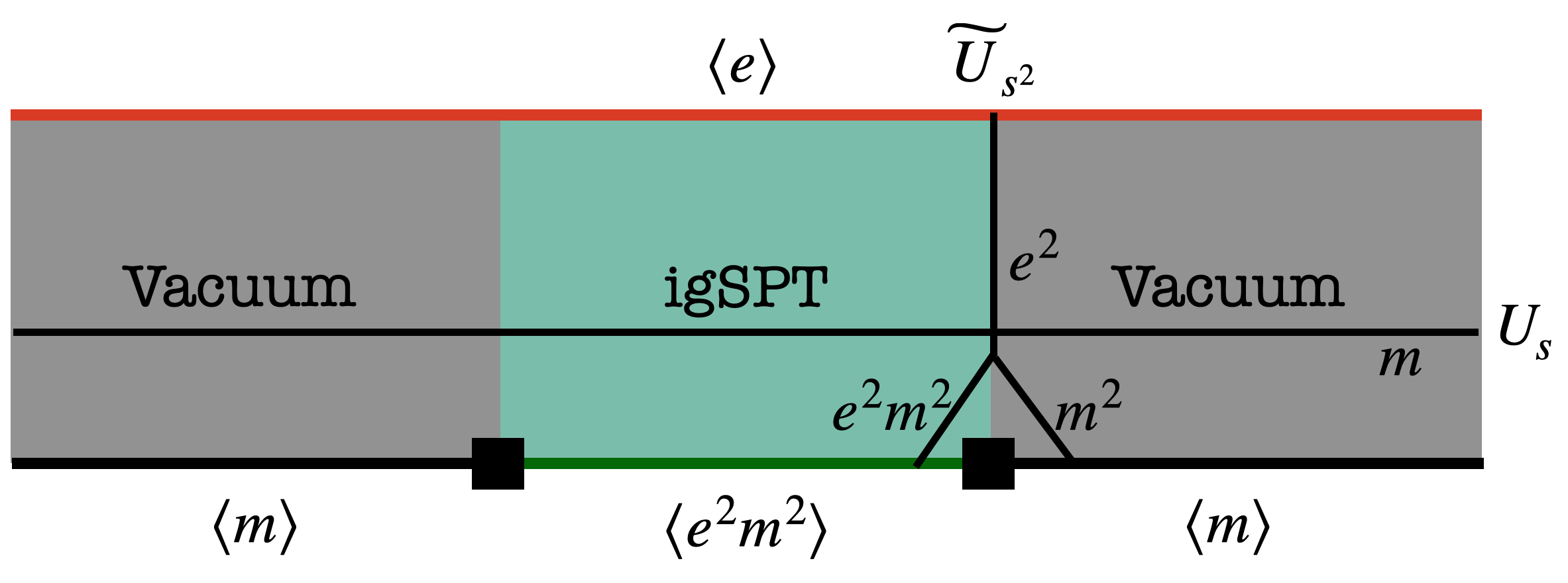}
    \caption{\textbf{Edge modes of the $\bZ_4$-igSPT in the thin slab construction.}The slab contains a segment of igSPT with two interfaces to the trivial symmetric gapped phase (vacuum). The $\bZ_2^A$ action can be localized to the interfaces, similar to the localization of symmetry actions of SPTs. The localized $A$-symmetry action intersects the $G$-symmetry action ($m$-string) at one point. Since $e^2$ and $m$ has mutual $\pi$ statistics, the two operators anti-commute, giving rise to a 2-fold GSD.}
    \label{fig:igSPT1}
    \end{figure*}

The $\bZ_4$-igSPT is a gapless system with a non-anomalous symmetry $\bZ_4$\footnote{The $\bZ_2^G$ anomaly of this igSPT is only emergent and the full system has no $\bZ_4$ anomaly.}. Thus  we expect that it is dual to a symmetric, partially-confining condensation of $D(\bZ_4)$. A condensation that satisfies these conditions is $\A=\langle e^2m^2\rangle$. Let us analyze this condensation. After condensing $e^2m^2$, the deconfined anyons are generated by $e^2,em$. $e^2$ and $em$ both have order 2 after condensing $e^2m^2$: $(e^2)^2=1$ trivially, and $(em)^2$ is the condensed anyon, which is identified as vacuum. Thus we have in total 4 inequivalent anyons after the condensation: $1,e^2,em,e^3m$. Interestingly, both $em$ and $e^3m$ are not bosons: $em$ is a semion with $\theta_{em}=i$ and $e^3m$ is an anti-semion with $\theta_{e^3m}=-i$. We conclude that the post-condensation topological order is the twisted quantum double $D_\omega(\bZ_2)$, also known as the double semion theory. The twist $\omega\in \Z^3[\bZ_2,U(1)]$ is the Levin-Gu anomaly. 

By invoking the LLEP, we see that the $e^2m^2$ condensation describes a 1+1D $\bZ_4$ symmetric gapless system whose local low energy properties are the same as that of a system with $\bZ_2$ symmetry and the Levin-Gu anomaly. This means the dual 1+1D system has a low energy sector with an effective anomalous $\bZ_2$ symmetry. This is exactly the structure of the $\bZ_4$-igSPT, therefore it is tempting to identify the condensation $\A=\{1,e^2m^2\}$ with the $\bZ_4$-igSPT. How does one describe properties of the $\bZ_4$-igSPT other than the emergent anomaly in the SymTFT language? Using the thin-slab construction, we show below that all physical properties of the $\bZ_4$-igSPT can be extracted from the $e^2m^2$ condensation.
\subsubsection{String order parameter from SymTFT}
An important feature of a non-trivial gSPT is the existence of non-trivial string order parameters. The charges carried by ends of these string order parameters can be used to distinguish different gSPT phases. In the $\bZ_4$-igSPT example, this can be seen by looking at the following string operator: $\S_{a,b}:=\sigma_a^z\prod_{i=a}^{b-1}e^{\frac{i\pi}{2}(1-\tau^x_{i+1/2})}\sigma^z_b$. In the low energy space below $\Delta$, we have $\langle \S_{a,b}\rangle=\langle \sigma^z_a\sigma_a^z\sigma^z_b\sigma^z_b\rangle=1$. Therefore this string operator acquires non-zero vacuum expectation value. Notice the end point of this string operator is charged under the $U_s$ symmetry: $U_s \S_{-\infty,b}U_s^{\dagger}=-\S_{-\infty,b}$. Now let us use the thin-slab construction with the $e^2m^2$-condensation to reproduce this result. Following the standard prescription, we put the $\A^e=\{1,e,e^2,e^3\}$ condensation on the top boundary and the $\A^{igSPT}=\{1,e^2m^2\}$ condensation on the bottom boundary. As we show in figure Fig.~\ref{fig:igSPT_string}, a non-local $H$-shaped operator can then be constructed in the thin-slab. In the middle of this operator is an $m^2$-string, while near the end of it the $m^2$ string joins an $e^2$ string from the top boundary to form an $e^2m^2$ string which is then absorbed by the bottom boundary. The structure of this operator mimics that of the operator $\S_{a,b}$ constructed before. Specifically, the end of this operator anti-commutes with an $m$ string that intersects it, see Fig~\ref{fig:igSPT_string}. Since this operator does not create any excitations, it acts as identity on the physical state represented by the thin-slab. Therefore it is a non-local string order parameter that has non-zero vacuum expectation value, with ends charged under the $U_s$ symmetry. We see that by utilizing the thin-slab construction, one can reconstruct the string order parameters of a gSPT from its dual anyon condensation.
\subsubsection{Edge modes from SymTFT}
Another important property of non-trivial gSPTs is the edge modes that appear when the system is put on an open chain. To observe the edge modes of the $\bZ_4$-igSPT in the thin-slab construction, consider a setup where the igSPT thin-slab is put in adjacent with slabs that represent the vacuum. The vacuum is represented as a all-flux-condensed boundary. This can be seen from, for example, the fact that condensing all fluxes on the bottom boundary gives a symmetric gapped 1+1D system with no non-trivial string order parameter. Therefore in this setup we have on the bottom boundary a segment of $e^2m^2$-condensation that is adjacent to $m$-condensations.  As we show in Fig.~\ref{fig:igSPT1}, the symmetry action by $\bZ_2^A$, i.e. a horizontal $m^2$-string, can be localized to the interfaces between the igSPT and the vacuum, and the localized $\bZ_2^A$ symmetry action anti-commutes with the symmetry action by $\bZ_2^G$, i.e. a horizontal $m$-string. This is exactly the algebra of symmetries of the $\bZ_4$ igSPT: The edge of the gapped symmetry $\widetilde{U}_{s^2}$ is charged under the gapless symmetry $U_s$.
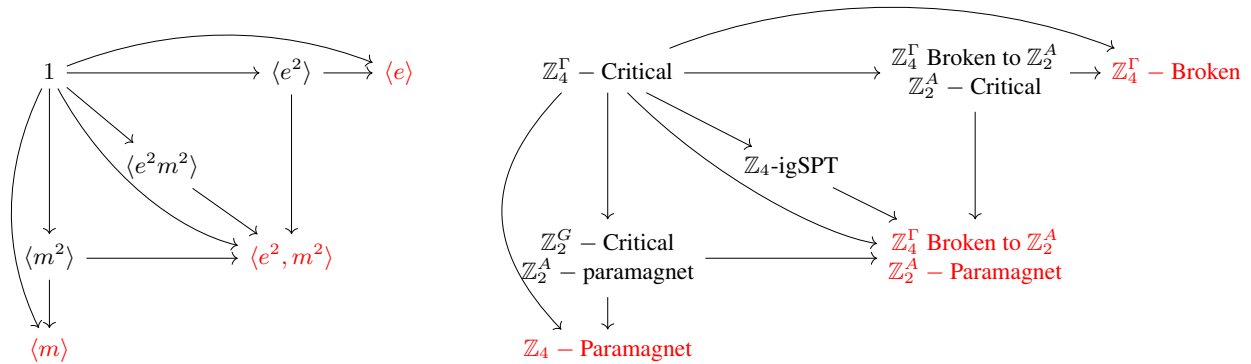
\begin{figure*}
    \[\begin{tikzcd}[ampersand replacement=\&,column sep=small,row sep=scriptsize]
        1 \&\& {\langle e^2\rangle} \& {\textcolor{red}{\langle e\rangle}} \&\&\& {\mathbb{Z}_4^\Gamma-\text{Critical}} \&\& {\begin{matrix} \mathbb{Z}^\Gamma_4~\text{Broken to}~\mathbb{Z}^A_2 \\ \mathbb{Z}^A_2-\text{Critical} \end{matrix}} \& {\textcolor{red}{\mathbb{Z}^\Gamma_4-\text{Broken}}} \\
        \& {\langle e^2m^2\rangle} \&\&\&\&\&\& {\mathbb{Z}_4\text{-igSPT}} \\
        {\langle m^2\rangle} \&\& {\textcolor{red}{\langle e^2,m^2\rangle}} \&\&\&\& {\begin{matrix}\mathbb{Z}_2^G-\text{Critical}\\\bZ_2^A-\text{paramagnet}\end{matrix} }\&\& {\begin{matrix}\textcolor{red}{\mathbb{Z}_4^\Gamma~\text{Broken to}~\mathbb{Z}^A_2}\\\textcolor{red}{\mathbb{Z}^A_2-\text{Paramagnet}}\end{matrix}} \\
        {\textcolor{red}{\langle m\rangle}} \&\&\&\&\&\& {\textcolor{red}{\mathbb{Z}_4-\text{Paramagnet}}}
        \arrow[from=3-7, to=4-7]
        \arrow[from=3-7, to=3-9]
        \arrow[from=1-9, to=3-9]
        \arrow[from=1-9, to=1-10]
        \arrow[from=2-8, to=3-9]
        \arrow[from=1-7, to=3-7]
        \arrow[from=1-7, to=2-8]
        \arrow[from=1-7, to=1-9]
        \arrow[from=1-3, to=1-4]
        \arrow[curve={height=-18pt}, from=1-1, to=1-4]
        \arrow[from=1-1, to=1-3]
        \arrow[from=1-1, to=3-1]
        \arrow[curve={height=18pt}, from=1-1, to=4-1]
        \arrow[curve={height=18pt}, from=1-1, to=3-3]
        \arrow[from=1-1, to=2-2]
        \arrow[from=2-2, to=3-3]
        \arrow[from=1-3, to=3-3]
        \arrow[from=3-1, to=3-3]
        \arrow[from=3-1, to=4-1]
        \arrow[shift right=5, curve={height=30pt}, from=1-7, to=4-7]
        \arrow[curve={height=-30pt}, from=1-7, to=1-10]
        \arrow[curve={height=18pt}, from=1-7, to=3-9]
    \end{tikzcd}\]
    \caption{\textbf{Stability of the $\bZ_4$-igSPT from SymTFT} On the left we draw a diagram where vertices are condensations of $D(\bZ_4)$, and there is an arrow between two condensations if one is a subset of the other, the arrow is pointing in the direction of the larger set. The condensations in red are fully-confining. On the right we draw the same diagram but with vertices replaced by the corresonding 1+1D phases, the phases in red are gapped. The $\bZ_4$-igSPT is only adjacent to the $\bZ_4$-Ising critical point and the partially symmetry broken phase, therefore the gaplessness of the igSPT is protected.}
    \label{fig: dz4 condense}
\end{figure*}
\subsubsection{Stability of the igSPT from SymTFT}
Starting with a 1+1D phase dual to a condensation $\A$, going to an adjacent phase amounts to adding anyons to the list $\A$ or deleting some anyons from $\A$~\cite{Chatterjee_2023_2}. This relation constrains the phase diagram of the dual 1+1D system. Let us now list all possible condensations of the $\bZ_4$ Toric code and their relations, shown in Fig.~\ref{fig: dz4 condense}. In this diagram, we connect two condensations if one is a subset of the other. It is clear from the diagram that the $\bZ_4$-igSPT is stable: although the system has no $\bZ_4$-anomaly, in order to go to the gapped $\bZ_4$-symmetric phases($\bZ_4$ paramagnet) the system has to either go through the $\bZ_4$-critical point where the gap to the $A$-dof is closed, or open up a gap for the $G$-dof by breaking the $\bZ_2^G$ symmetry and then restore the symmetry by condensing domian walls of $\bZ_2^G$ and going through a $\bZ_2^G$-critical point. From the anyon condensation perspective, the anyon $m$ can not be added to the condensation $\{1,e^2m^2\}$ because it braids non-trivially with $e^2m^2$, this indicates a direct transition from $\bZ_4$-igSPT to $\bZ_4$-paramagnet is impossible. This is an instance of ``anomaly protected gaplessness'' that appears in all igSPTs, where the existence of an emergent anomaly gives obstruction to directly gapping out the system while preserving the symmetry. We see that the relation between different anyon condensation patterns provides a SymTFT description of this mechanism. 

The $\bZ_4$-igSPT example demonstrates that the SymTFT is capable of providing comprehensive characterization of 1+1D gSPTs. Therefore it is plausible that every $\Gamma$-gSPT is dual to certain symmetric, partially-confining condensation of the quantum double $D(\Gamma)$. However in order to fully establish such a duality it is crucial to understand the classification of two types of objects in two categories: on one hand the classification of gSPTs in 1+1D, and on the other hand the classification of symmetric partially confining condensations of quantum doubles. This will be the goal of the rest of this work. After carefully examining the structure of 1+1D gSPTs, we provide a classification of 1+1D gSPTs protected by internal unitary finite symmetry. The classification itself turns out to provide us some surprising  insights into the structure of 1+1D gSPTs. For example, we find that there is an edge/bulk (0+1D/1+1D) correspondence for 1+1D gSPTs in a strict sense to be defined. After having a complete characterization of 1+1D gSPTs, we review the established theory of anyon condensation and discuss the structure of symmetric partially confining condensations of 2+1D quantum doubles as well as its connection to 1+1D gSPTs. A 1-1 correspondence between the two will by then become evident.
\section{Structure of 1+1D bosonic gSPTs\label{sec:gspt_structure}}
In order to establish formally the connection between 1+1D gSPTs and their SymTFT dual, it is crucial to have a complete understanding of the structure of 1+1D gSPTs. This is the task of this section. 
\subsection{Algebra of symmetry and consistency conditions}
Let us consider a general 1+1D gSPT with a gapped sector and a gapless sector having the structure of a symmetry extension
\begin{align}
    1\rightarrow A\xrightarrow[]{j} \Gamma\xrightarrow[]{p} G\rightarrow 1.\label{eq: groupext}
\end{align}
Here $A\trianglelefteq \Gamma$ is a normal subgroup of $\Gamma$, not necessarily Abelian. Being a normal subgroup, we have $\gamma A \gamma^{-1}=A$ for any $\gamma\in\Gamma$, and this defines a $\Gamma$-action on $A$.
We ask what extra data besides the group extension is needed to specify a gSPT. From our experience with SPTs, we know that the low energy physics of an SPT is completely specified by the anomalous symmetry action on the boundary. For a 1+1D SPT, the algebra of edge symmetry action, namely the projective representation of edge symmetry actions, is the origin of the SPT edge modes. Motivated by this, we look at symmetry actions and their algebra in a 1+1D gSPT,  this algebra will similarly determine the edge modes of the gSPT.  First of all the UV $A$-symmetry actions act trivially in the bulk in the low energy sector, since by definition there is no $A$-charges in the gapless sector in the bulk. Therefore on a closed chain we expect to have $U^{\text{IR}}_a=1$. In the following we will drop the IR superscript and it is understood that operators are restricted to the gapless low energy subspace. However on an open chain the $A$-action can still act non-trivially on the edge of the chain, this is known as symmetry fractionalization principle for symmetries acting on gapped degrees of freedom~\cite{Pollmann_2010,Turner_2011,Verresen_2017}. Therefore an $A$-symmetry action $U_a$ reduces to an edge action $U_a=U_a^L\otimes U_a^R$ below the gap to $A$-dof. Similar to the edge symmetry action of SPTs, the localized edge actions $\widetilde{U}_a:=U_a^{L}$ only need to form a projective representation of the gapped symmetry group $A$: $\widetilde{U}_a\widetilde{U}_b=\eta(a,b)\widetilde{U}_{ab}$, where $\eta\in \Z^2[A,U(1)]$ is a 2-cocycle of $A$. We also need to consider the interplay between the localized edge action $\widetilde{U}_a$ and the non-localized $G$-action. Consider a general $\gamma\in\Gamma$ action, and the conjugation $U_\gamma \widetilde{U}_a U_\gamma^{-1}$, the group law states that $U_\gamma U_a U_\gamma^{-1}=U_{\gamma a \gamma^{-1}}$, however using the decomposition $U_a=U^L_a\otimes U^R_a$, we have 
\begin{align}
    U_\gamma U_a^L U_{\gamma}^{-1}\otimes U_\gamma U^R_a U_\gamma^{-1}=U^L_{\gamma a \gamma^{-1}}\otimes U^R_{\gamma a\gamma^{-1}}.\label{eq:udec}
\end{align}
Notice that $U_\gamma U_a^{L,R} U_{\gamma}^{-1}$ is an operator supported on the left/right edge: away from the edge $U^{L,R}_a$ acts as identity and $U_\gamma U_a^{L,R} U_{\gamma}^{-1}=U_\gamma U_\gamma^{-1}=1$. Therefore both sides of~\eqref{eq:udec} decompose into tensor products of operators acting on separate Hilbert spaces, thus each factor must be equal up to a phase:
\begin{align}
    U_\gamma \widetilde{U}_a U_\gamma^{-1}=\epsilon(a,\gamma)\widetilde{U}_{\gamma a\gamma^{-1}}\label{eq:projact}
\end{align}
The phase $\epsilon(a,\gamma)$ can be thought of as measuring the charges of the fractionalized $\widetilde{U}_a$ action under $U_\gamma$. The data $(\eta,\epsilon)$ naturally satisfies a set of consistency conditions. For example, $U_\gamma(\widetilde{U}_a\widetilde{U}_b)U_\gamma^{-1}$ should be equal to $(U_\gamma\widetilde{U}_aU_\gamma^{-1})\cdot (U_\gamma\widetilde{U}_bU_\gamma^{-1})$, which leads to an equation
\begin{align}
    \eta(a,b)\epsilon(ab,\gamma)=\eta(\gamma a\gamma^{-1},\gamma b \gamma^{-1})\epsilon(a,\gamma)\epsilon(b,\gamma).
\end{align}  
We perform careful analysis of consistency conditions in appendix~\ref{app: consistency conditions} and summarize them here:
\begin{figure}
        \includegraphics[width=1.\columnwidth]{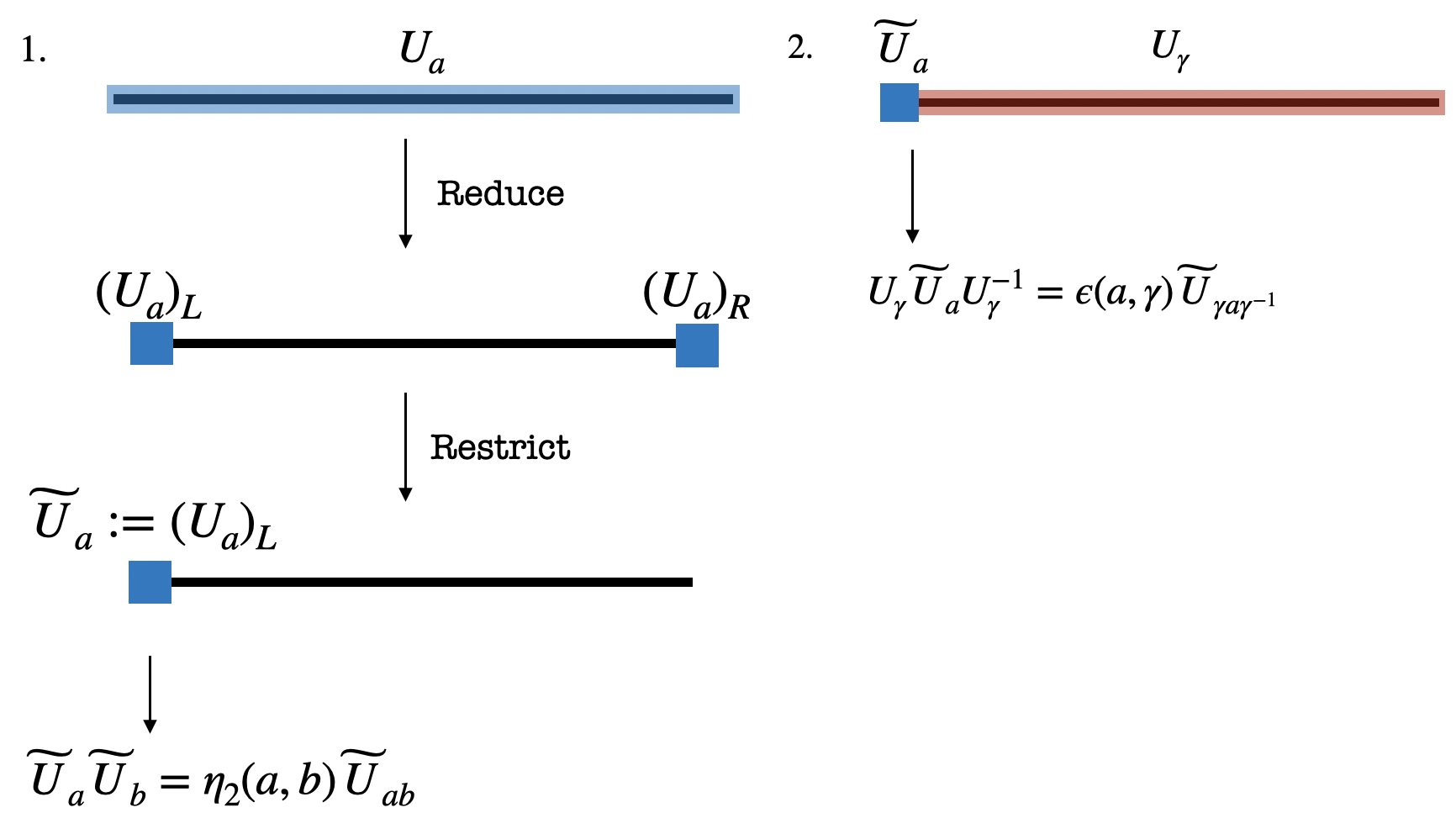}
        \caption{\textbf{The algebra of symmetry action of a gSPT.} 1. UV symmetry actions $U_a, a\in A$ reduce to edge actions in the gapless sector, and the restriction of them to one of the endpoints make up a projective representation of $A$ given by a 2-cocycle $\eta$. 2. The restriction of an $A$-symmetry action to one of the endpoints fails to satisfy the commutation relation with another symmetry action.}
        \label{fig:gSPT_algebra}
\end{figure}

\begin{mdframed}
    A 1+1D $\Gamma$-gSPT is defined by $(A\lhd \Gamma, \eta,\epsilon)$, where $A$ is a central subgroup of $\Gamma$, and $\eta,\epsilon$ satisfy the following conditions:
    \begin{align}
    &\eta\in \Z^2[A,U(1)],~\epsilon: A\times \Gamma\to U(1)\nonumber\\
    &\frac{\epsilon(a,\gamma)\epsilon(b,\gamma)}{\epsilon(ab,\gamma)}=\frac{\eta(a,b)}{\eta(\gamma a\gamma^{-1},\gamma b \gamma^{-1})}\label{eq:consistency conditions 1}\\
    &\epsilon(a,\gamma_1\gamma_2)=\epsilon(\gamma_2 a\gamma_2^{-1},\gamma_1)\epsilon(a,\gamma_2)\label{eq:consistency conditions 2}\\
    &\epsilon(a,b)=\frac{\eta(b,a)}{\eta(bab^{-1},b)},~\forall a,b\in A \label{eq:consistency conditions 3}
\end{align}
\end{mdframed}
Moreover, the decomposition $U_a=U^L_a\otimes U^R_a$ only defines the localized $A$-action $U^{L,R}_a$ up to a phase. Redefining $U^L_a\to \alpha(a)U^L_a, U^R_a\to \alpha^*(a)U^R_a$ does not change the overall symmetry action $U_a$. This will have the effect of modifying the 2-cocycle $\eta$ by a coboundary, and will also affect the $\epsilon$ factor by $\epsilon(a,\gamma)\to \epsilon(a,\gamma)\frac{\alpha(\gamma a\gamma^{-1})}{\alpha(a)}$. Since the choice of the phase factor $\alpha$ is artifitial, two solutions to the consistency conditions related by such a transformation should be viewed as defining the same gSPT.
\begin{mdframed}
    Two sets of data $(\eta,\epsilon),~(\eta',\epsilon')$ define the same gSPT, if they are related by
    \begin{align}
        &\eta'(a,b)=\eta(a,b)\frac{\alpha(a)\alpha(b)}{\alpha(ab)}\\
        &\epsilon'(a,\gamma)=\epsilon(a,\gamma)\frac{\alpha(\gamma a\gamma^{-1})}{\alpha(a)}.
    \end{align}
\end{mdframed}
We will refer to this equivalence relation as gauge equivalence, and a transformation of a pair $(\eta,\epsilon)$ by the equivalence relation as a gauge transformation.The set of equivalent classes of solutions $(\eta,\epsilon)$ to the consistency conditions is our classification of 1+1D bosonic gSPT with unitary symmetry.
 
\paragraph{Finiteness of the classification.} It is not immediately clear that the classification above is finite or not, although in many cases it can be verified by hand that the solutions are finite modulo gauge transformation. However we will show that gSPTs are in 1-1 correspondence with symmetric partially confining condensations of 2+1D quantum doubles, and the latter always has a finite classification for any finite gauge group. Thus the SymTFT dual will provide a definite answer to the finiteness question, at least for finite $\Gamma$. Thus there are only finite many gSPTs in 1+1D for any finite symmetry group $\Gamma$. A seemingly contradiction is that we can always stack a gSPT with a CFT that transforms trivially under the symmetry without changing the data $(\eta,\epsilon)$. The resolution is that we should view the classification as kinematic and does not determine what CFT lives in the bulk. That is, it classifies possible $\Gamma$-symmetry enriching patterns of all $G$-symmetric gapless states. Alternatively, one can take the bulk phase as an extra data that needs to be specified besides $(\eta,\epsilon)$. However the data $(\eta,\epsilon)$ does put constraints on the possible gapless state. For example, the gapless sector needs to have \textbf{nontrivial} $G$-symmetry, since if $G$-symmetry acts trivially on the gapless sector, it would have been identified as a gapped symmetry. Also, as we will show soon, the pair $(\eta,\epsilon)$ determines the emergent anomaly of the bulk, which also needs to be matched by the gapless sector.
\paragraph{Group structure of the classification.} There is an obvious group structure of the classification. Namely since the consistency conditions are all multiplicative in $\eta, \epsilon$, the product of two solutions is again a solution. The solution $\eta=1,\epsilon=1$ serves as group identity, and the inverse of an element $(\eta,\epsilon)$ is $(\eta^{-1},\epsilon^{-1})$. It is tempting to interpret the group structure as the physical operation of stacking two gSPTs together, however there is a subtlty. For example, if we stack a gSPT with a CFT$^A$ bulk and another with a CFT$^B$ bulk, then the stacked system will have CFT$^A\times$CFT$^B$ bulk. Therefore the stacking operation only defines a group structure modulo the bulk state. This is tight to  the fact that the classification is really a classification of symmetry enriching patterns.
\paragraph{Invariants of gSPTs.} The data $\eta,\epsilon$ provides some discrete invariants for a gSPT which will be useful in practice. Consider a given $a\in A$, let $C_a(\Gamma)$ be the centralizer of $a$ in $\Gamma$, i.e. the subgroup consisting of elements that commute with $a$. Then we have $U_c \widetilde{U}_aU_c^\dagger=\epsilon(a,c)\widetilde{U}_a,~\forall c\in C_a(\Gamma)$, and the factor $\epsilon(a,c)$ is now gauge invariant: a gauge transformation changes it to $\epsilon'(a,c)=\epsilon(a,c)\frac{\alpha(caa^{-1})}{\alpha(a)}=\epsilon(a,c)$.  Therefore $\epsilon$ restricted to $\{a\}\times C_a(\Gamma)$ provides an invariant for any $a\in A$. This is nothing but the charge carried by $\widetilde{U}_a$ under $C_a(\Gamma)$. In general $\widetilde{U}_a$ does not have a well-defined $\Gamma$-charge, since a $U_\gamma$ conjugation maps it to $\widetilde{U}_{\gamma a\gamma^{-1}}$ up to a phase factor. Restricting to $C_a(\Gamma)$ gives a well-defined charge of $\widetilde{U}_a$. However, we should stress that in general these invariants do not determine the gSPT completely. The method of using charges of edge symmetry actions as invariants also applies to 1+1D SPTs, and it is known that there are examples where such charges do not fully determine the SPT (although the minimal counter-example requires a symmetry group of order 128)~\cite{Pollmann_2012}. Put the counter-example as the gapped sector will give a counter-example for gSPTs.

\subsection{Emergent anomaly of gSPT and a bulk-edge (1+1D/0+1-D) correspondence}
We next show that the data $(\eta,\epsilon)$ determines the emergent anomaly of the gSPT. Since the data $(\eta,\epsilon)$ are directly related to the edge modes of the gSPT, the result that they determine the emergent anomaly can then be viewed as an edge-to-bulk map for gSPTs: the edge modes determine the emergent anomaly. What's interesting is that the converse is also (partially) true: the emergent anomaly determines $(\eta,\epsilon)$ up to stacking with a $\Gamma$-SPT. Therefore there is a bulk-edge correspondence\footnote{not to be confused with the holographic duality between 1+1D gSPTs and 2+1D quantum doubles.} for gSPTs in the following sense: the edge modes determine the bulk emergent anomaly, and the bulk emergent anomaly determines the edge modes up to stacking with an SPT.

There are more than one way to detect the anomaly of a symmetry. For example anomaly amounts to obstruction to consistently coupling the system to a background gauge field, equivalently anomaly is the obstruction to consistently inserting a defect network of the symmetry. A practical way of computing the anomaly when the symmetry action is explicitly known is provided by~\cite{Else_2014}, where anomaly is formulated as the obstruction to implementing the symmetry on a system with boundary. Here we review this approach briefly. 

For a symmetry action $U_g,~g\in G$ acting on a 1+1D system with open boundary condition, if we compare $U_gU_h$ and $U_{gh}$, they may fail to be equal up to operators supported on the edge: 
\begin{align}
    U_gU_h=\Omega(g,h)U_{gh}, \label{eq: anomaly1}
\end{align}
where $\Omega(g,h)$ is a local unitary operator supported on the  edge: $\Omega(g,h)=\Omega^L(g,h)\otimes \Omega^R(g,h)$. From the definition~\eqref{eq: anomaly1} the operators $\Omega$ must satisfy a condition from the associativity of the $U_g$ operators,
\begin{align}
    &(U_gU_h)U_k=U_g(U_hU_k)\nonumber\\
    &\Rightarrow \Omega(g,h)\Omega(gh,k)=U_g\Omega(h,k)U_g^{-1} \Omega(g,hk).
\end{align}
However, since the operator $\Omega$ splits into tensor product of $\Omega^L$ and $\Omega^R$, the restriction of it to one of the edges only need to satisfy the consistency condition up to a phase:
\begin{align}
    \omega_3(g,h,k)\Omega^L(g,h)\Omega^L(gh,k)=U_g\Omega^L(h,k)U_g^{-1} \Omega^L(g,hk). \label{eq: anomaly2}
\end{align}
The phase $\omega_3$ can be shown to be a 3-cocycle and represents the anomaly of the symmetry action $U_g$. 

We now show that the data $\eta,\epsilon$ suffice to determine the emergent anomaly of a gSPT. The group extension~\eqref{eq: groupext} defines a 2-cocycle of $G$ with $A$ coefficients: $e_2\in Z^2[G,A]$, with $G$ acting on $A$ by conjugation. The role of this 2-cocycle is to specify the group law of $\Gamma$ from that of $G$ and $A$. One can write elements of $\Gamma$ as pairs $\gamma=(a,g),a\in A,g\in G$. The group law of $\Gamma$ is then given by $(a,g)\cdot (b,h)=(a\cdot gbg^{-1}\cdot e_2(g,h),gh)$. Now with this understanding of group structure of $\Gamma$, one can compute the composition of symmetry actions by $G$: 
\begin{align}
    U_gU_h=U_{(e_2(gh),gh)}=U_{e_2(g,h)}U_{gh}.
\end{align}
The operator $U_{e_2(g,h)}$ when restricted to the gapless sector acts only non-trivially on the edge, therefore the $G$-action(when restricted to the gapless sector) does take the form of \eqref{eq: anomaly1}, with $\Omega(g,h)=U_{e_2(g,h)}$ and $\Omega^{L}(g,h)=\widetilde{U}_{e_2(g,h)}$. Now we can calculate the anomaly according to equation~\eqref{eq: anomaly2}:
\begin{align}
    &\Omega^L(g,h)\Omega^L(gh,k)=\widetilde{U}_{e_2(g,h)}\widetilde{U}_{e_2(gh,k)}\nonumber\\
    &=\eta(e_2(g,h),e_2(gh,k))\widetilde{U}_{e_2(g,h)e_2(gh,k)}, \label{eq: anomaly3}\\ 
    &U_g\Omega^L(h,k)U_g^{-1}\Omega^L(g,hk)= U_g \widetilde{U}_{e_2(h,k)}U_g^{-1}\widetilde{U}_{e_2(g,hk)}\nonumber\\
    &=\epsilon(e_2(h,k),g)\widetilde{U}_{g e_2(h,k)g^{-1}}\widetilde{U}_{e_2(g,hk)}\nonumber\\
    &=\epsilon(e_2(h,k),g)\eta(g e_2(h,k)g^{-1},e_2(g,hk))\widetilde{U}_{g e_2(h,k)g^{-1}e_2(g,hk)}\label{eq: anomaly4}
\end{align}
The cocycle condition on $e_2$ is 
\begin{align}
    ge_2(h,k)g^{-1}e_2(g,hk)=e_2(g,h)e_2(gh,k),
\end{align} 
therefore the last lines of \eqref{eq: anomaly3} and \eqref{eq: anomaly4} only differ by a phase factor, which is the anomaly 3-cocycle of the $G$-symmetry action:
\begin{mdframed}
    \begin{align}
        \omega_3(g,h,k)=\frac{\eta(ge_2(h,k)g^{-1},e_2(g,hk))}{\eta(e_2(g,h),e_2(gh,k))}\epsilon(e_2(h,k),g)\label{eq:emergent anomaly}
    \end{align}
\end{mdframed}

\paragraph{Trivial vs. nontrivial extension.} From the expression~\eqref{eq:emergent anomaly} it is clear that if the group extension of a gSPT is trivial, then $e_2=1$ and the emergent anomaly vanishes. Thus in order to have an emergent anomaly the group extension must be non-trivial, i.e. the full symmetry group $\Gamma$ can not be a direct product or semi-direct product of the gapped symmetry $A$ and the gapless symmetry $G$.

\paragraph{gSPTs of SPT-pump type.}
Any solution to the consistency conditions~\eqref{eq:consistency conditions 1}-\eqref{eq:consistency conditions 3} defines a gSPT and determines the emergent anomaly through \eqref{eq:emergent anomaly}. Consider a special case where $\eta=1$ and the group extension is central. In this case elements of $A$ commute with all elements of $\Gamma$: $\gamma a\gamma^{-1}=a$, and the consistency conditions on $\epsilon$ reduce to 
\begin{align}
    &\epsilon(a,\gamma)\epsilon(b,\gamma)=\epsilon(ab,\gamma)\\
    &\epsilon(a,\gamma_1\gamma_2)=\epsilon(a,\gamma_1)\epsilon(a,\gamma_2)\\
    &\epsilon(a,b)=1~\forall a,b\in A.
\end{align}
The first two conditions state that $\epsilon$ is now linear in both factors, and the last condition states that the $\epsilon$ is well-defined on the quotient $A\times \Gamma/A=A\times G$. Viewed from the original definition, $U_\gamma \widetilde{U}_a U_\gamma^{-1}=\epsilon(a,\gamma)\widetilde{U}_a$, these conditions imply that now every $a\in A$ action carries a well-defined charge under $G$. The anomaly according to~\eqref{eq:emergent anomaly} is $\omega_3(g,h,k)=\epsilon(e_2(h,k),g)=\epsilon\cup e_2(g,h,k)$, and the edge modes of the gSPT can be intuitively understood as $A$-actions pumping $G$-charges (which are 0+1-D SPTs of $G$). Such gSPTs have been studied in~\cite{Wen_2023} and the SPT-pump mechanism was also used to understand igSPTs in higher dimensions.
\paragraph{Bulk-edge (1+1D/0+1-D) correspondence for gSPT}
Interestingly the emergent anomaly determines the data $(\eta,\epsilon)$ up to stacking with $\Gamma$-SPTs. This means up to $\Gamma$-SPTs the physics of a $\Gamma$-gSPT is completely determined by its emergent anomaly. This has an immediate consequence that the two definitions of igSPT are equivalent: 1. A gSPT is intrinsic if the edge modes can not be trivialized by stacking with an SPT. 2. A gSPT is intrinsic if it has a nontrivial emergent anomaly. We provide a comprehensive derivation of these statements in appendix~\ref{app: gSPTbulkedge}. The idea of the derivation can be summarized as follows. Denote the group extension as 
\begin{align}
    1\rightarrow A\xrightarrow[]{j} \Gamma\xrightarrow[]{p} G\rightarrow 1\label{eq:groupext}
\end{align}
Since the anomaly is trivialized upon pull-back by $p$, we have on the cochain level $p^*(\omega)=d(\alpha)$, where $\alpha\in \C^{2}[\Gamma,U(1)]$ is a 2-cochain of $\Gamma$. The composition of maps $p\circ i$ is trivial since the sequence~\ref{eq:groupext} is exact: $p\circ j=0$. This implies the composition of pull-backs $(p\circ j)^*=j^*\circ p^*=0$ is also trivial. Therfore $0=j^*\circ p^*(\omega)=j^*(d\alpha)=d(j^*\alpha)$, meaning that $j^*\alpha$ is a 2-cocycle of $A$ representing a 1+1D $A$-SPT. This is to be identified with the SPT of the gapped sector, i.e. $\eta=j^*\alpha$. $\epsilon$ can be found by considering the value $\alpha((a,1),(1,g))$. Since the trivializing cochain $\alpha$ is not unique, this procedure only determines $\eta,\epsilon$ up to stacking with $\Gamma$-SPTs.  In section~\ref{sec:SPT_stack}  we will address the question of stacking gSPTs with SPTs.

\section{Condensation of quantum double and gSPT\label{sec:general_condense}}
The $\bZ_4$-igSPT example suggests that symmetric partially-confining condensations in the 2+1D quantum double $\D(\Gamma)$ correspond to $\Gamma$-gSPTs. Now that we have a classification of 1+1D gSPTs, we are in position to establish this correspondence in a rigorous manner. We first briefly summarize the results.
\subsection{Summary of results}
 The SymTFT for a symmetry $\Gamma$ is the quantum double $D(\Gamma)$, also known as the gauge theory with gauge group $\Gamma$. Anyons of $D(\Gamma)$ are gauge charges, gauge fluxes and composites of charge and fluxes (dyons). An anyon condensation amounts to specifying a set of condensed anyons and also a fusion channel for fusing any two condensed anyons. For quantum doubles, an anyon condensation can be specified as follows. One can first separate the condensation into two steps. In the first step we only condense pure charges, which will lead to gauge symmetry breaking from $\Gamma$ to some subgroup $H<\Gamma$. This is dual to a global symmetry breaking in the SymTFT setup, therefore for describing symmetric 1+1D systems one can neglect this step and set $H=\Gamma$. In the second step we only condense fluxes and/or dyons. In general a dyon has a flux sector and a charge sector. The contents of the condensed dyons and their fusion channels are then specified by 3 objects: a normal subgroup $A< \Gamma$ and two phase factors $\epsilon: A\times \Gamma\to U(1), \eta: \Gamma\times \Gamma\to U(1)$. $A$ specifies the flux sectors of the condensed dyons, $\epsilon$ determines the charge sector of each condensed dyon, and $\eta$ determines the fusion phase of fusing two condensed dyons. In general fluxes take value in $\Gamma$. Then the condensed dyons must have flux sectors taking values in the normal subgroup $A$. For a given dyon with flux sector $a\in A$, $\epsilon(a,\gamma)$ then measures the charge sector. It can be thought of as the phase that is picked up by braiding a flux of value $\gamma$ around the dyon with flux sector $a$. By collecting the phases $\epsilon(a,\gamma)$ resulting from braiding with all the fluxes one can then determine the charge sector of the dyon. Finally $\eta(a,b)$ is the fusion phase of fusing two condensed dyons with flux sectors $a\in A$ and $b\in A$. One immediately sees that the set of data needed to specify a symmetric condensation is identical to that is needed to specify a 1+1D $\Gamma$-gSPT. It can be shown that in order for the data $A,\epsilon,\eta$ to define a consistent boson condensation, they must satisfy certain consistency conditions. These turn out to be identical to the consistency conditions of 1+1D gSPTs~\eqref{eq:consistency conditions 1}-\eqref{eq:consistency conditions 3}. Moreover, a condensation gives rise to a new topological order. It can be shown that the condensation specified by the data $A,\epsilon,\eta$ is fully-confining if and only if $A=\Gamma$. And when $A\neq \Gamma$, the post-condensation topological order is in general a twisted quantum double of the quotient group $G:=\Gamma/A$: $D_\omega(G)$, and the twist $\omega$ turns out to be given by the same formula as~\eqref{eq:emergent anomaly}. This then establishes a 1-1 correspondence between symmetric partially-confining condensations of $D(\Gamma)$ and $\Gamma$-gSPTs.

In the rest of this section we review the formal theory of anyon condensation and discuss the structure of anyon condensation in 2+1D quantum doubles. 
\subsection{Review of condensation in 2+1D topological orders}
We first review aspects of UMTC relevant to our discussion and then discuss the description of anyon condensation in this language.
A general 2+1D topological order is described by a unitary modular tensor category(UMTC) $\EC$. Objects of $\EC$ correspond to anyons of the topological order. A UMTC is equipped with a tensor product of  pairs of objects $\otimes$, which describes the fusion of anyons, an associator $\alpha_{X,Y,Z}:(X\otimes Y)\otimes Z\to X\otimes(Y\otimes Z)$ that relates different orders of fusing three anyons, and a braiding $c_{X,Y}:X\otimes Y\to Y\otimes X$ that describes the process of exchanging two anyons. The data $\alpha_{X,Y,Z},c_{X,Y}$ satisfy certain consistency conditions, including the pentagon equation of the associators and the hexagon equation that relates the associator with the braiding~\cite{Kitaev_2006,bonderson2012non,kong2022invitation}.

A condensation amounts to bringing a large number of anyons in a region together to form a new phase from $\EC$. 
This physical process corresponds to obtaining a new UMTC from $\EC$ by specifying a condensable algebra $\A$ of $\EC$~\cite{kong2014anyon}. A condensable algebra describes the contents of condensed anyons, if several simple anyons $a_i$ are condensed, we can view $\A$ as the composite anyon $\A=a_1+a_2+\cdots$.  After condensing $\A$, we obtain a new phase $\ED$, with a new fusion rule $\otimes_\ED$. In the new phase $\A$ is identified as the vacuum, $\A=1_\ED$. The vacuum of the new phase must fuse trivially with itself: $\A\otimes_\ED\A=\A$. Since $\A$ is also an anyon of the original phase, and fusion channels of the new phase $\ED$ come from those of the original phase $\EC$, this means when viewed as an anyon of $\EC$, there must be a nontrivial fusion channel $\mu:\A\otimes_\EC \A\to \A$. This makes $\A$ an algebra in $\EC$. There are extra conditions on the algebra due to the physical requirements that the condensed anyons should be self and mutual bosons. These conditions make $\A$ a so called connected separable commutative algebra, or a condensable algebra for simplicity. Now consider an excitation $M$ that lives in the post-condensation phase $\ED$, the fusion $M\otimes_\ED \A$ is equal to $M$ since now $\A$ is the vacuum: $M\otimes_\ED \A=M\otimes_\ED 1_\ED=M$. On the other hand both $M$ and $A$ are also anyons(possibly composite ones) of the original phase, and fusion channels of $\ED$ come from that of the origin phase $\EC$. Therefore when viewed as anyons of $\EC$, there should be a nontrivial fusion channel $\nu_M: M\otimes_\EC \A\to M$. This makes $M$ a module over $\A$.  For example, in the toric code the composite anyon $m+f$ is a module over $1+e$, therefore is an excitation after condensing $1+e$. Deconfined anyons in the new phase must braid trivially with vacuum, then when viewed as anyons of the original phase, this means the diagram
\[\begin{tikzcd}[ampersand replacement=\&]
	{M\otimes_\EC \A} \&\& M \\
	\\
	{\A\otimes_\EC M} \&\& {M\otimes_\EC \A}
	\arrow["{\nu_M}", from=1-1, to=1-3]
	\arrow["{c_{M,\A}}"', from=1-1, to=3-1]
	\arrow["{c_{\A,M}}"', from=3-1, to=3-3]
	\arrow["{\nu_M}"', from=3-3, to=1-3]
\end{tikzcd}\]
commutes for any deconfined anyon represented by a module $M$. This condition makes $M$ a so called local module over $\A$. Thus the deconfined excitations in the new phase are given by local modules over $\A$ in the original phase. The category of local modules over $\A$, $\EC^{loc}_\A$ has a full UMTC structure induced from $\EC$~\cite{kong2014anyon}, therefore describes a valid topological order.

\subsection{Condensation of 2+1D quantum double}
Let us now focus on a specific kind of topological order of interests to us, the 2+1D quantum doubles. The quantum double $D(\Gamma)$ is a gauge theory with gauge group $\Gamma$. The anyons in $D(\Gamma)$ can generally be identified as gauge charges, gauge fluxes, and composites of charge and flux. Gauge fluxes are in general labelled by conjugation classes of the group $\Gamma$, and the charges are labelled by irreducible representations of the group. 
A unified way of describing anyons of $D(\Gamma)$ is to identify an anyon of $D(\Gamma)$ as a $\Gamma$-graded vector space $V=\oplus_{\gamma\in\Gamma}V_\gamma$, together with a $\Gamma$-action, such that $\gamma(V_{\gamma'})=V_{\gamma \gamma'\gamma^{-1}}$. 
In this way a charge corresponds to the case where the graded vector space is only supported at the identity element: $V=V_{1_\Gamma}$, then the anyon is completely determined by a representation of the group $\Gamma$, recovering the fact that charges are labelled by representations of $\Gamma$. On the other hand a flux corresponds to the case where the vector space is supported on at least one nontrivial element while the $\Gamma$-action is trivial. The vector space has to be supported on conjugation classes of $\Gamma$ to be invariant under the $\Gamma$-action, therefore a flux is labelled by a conjugation class of $\Gamma$. A generic bond state of charge and flux then corresponds to a graded vector space with nontrivial support as well as nontrivial $\Gamma$-action. 

Condensable algebras of $D(\Gamma)$ have been classified in~\cite{davydov2009modular}. A condensation in $D(\Gamma)$ can always be done in two steps. In the first step, only charges are condensed, reducing $D(\Gamma)$ to $D(H)$, for some subgroup $H<\Gamma$. This step can be thought of as gauge symmetry breaking from $\Gamma$ to $H$. The condensable algebra for this step is given by the algebra of functions over the coset $\Gamma/H$: $\bC(\Gamma/H)$. As an element of $D(\Gamma)$, $\bC(\Gamma/H)$ is a graded vector space supported only at the identify $1_\Gamma$, with the $\Gamma$-action defined by $(\gamma\cdot f)([\gamma']):=f([\gamma'\gamma^{-1}])$ for any function $f$ on the coset $\Gamma/H$. Clearly this defines a representation of $\Gamma$ and corresponds to a condensation of charges. Local modules over $\bC(\Gamma/H)$ can be shown to form a UMTC equivalent to the quantum double $D(H)$. In the second step no charges are condensed, and the condensable algebra is determined as follows:

\textit{(\cite{davydov2009modular}, Thm. 3.5.3)
A condensable algebra of $D(H)$ that does not condense any charges are determined by $\eta\in H^2(H,U(1))$, a normal subgroup $A\lhd H$, and a function $\epsilon:A\times H\to U(1)$, such that the following conditions are met:
\begin{align}
    &\frac{\epsilon(a,h)\epsilon(b,h)}{\epsilon(ab,h)}=\frac{\eta(a,b)}{\eta(h ah^{-1},h b h^{-1})}\label{eq:condense-1}\\
&\epsilon(a,h_1h_2)=\epsilon(h_2 ah_2^{-1},h_1)\epsilon(a,h_2)\label{eq:condense-2}\\
&\epsilon(a,b)=\frac{\eta(b,a)}{\eta(bab^{-1},b)},~\forall a,b\in A \label{eq:condense-3}
\end{align}
\label{prop-1}}
The algebra $\A[A,\eta,\epsilon]$ determined by these data can be described as follows. First $\A[A,\eta,\epsilon]$ should be an (composite) anyon of $D(H)$, i.e. an $H$-graded vector space with an $H$-action. As a graded vector space, $\A[A,\eta,\epsilon]=\oplus_{a\in A}\bC_a$ is supported on the subgroup $A$, let the basis of $\bC_a$ be $e_a$. The $H$-action is defined by $h(e_a):=\epsilon(a,h)e_{hah^{-1}}$. The algebra structure on $\A[A,\eta,\epsilon]$ is defined on the basis $e_a$ by $e_a\cdot e_b:=\eta(a,b)e_{ab}$. Requiring the algebra $\A[A,\eta,\epsilon]$ to be well-defined and condensable then leads to the consistency condition on $\epsilon$ and $\eta$ \eqref{eq:condense-1}-\eqref{eq:condense-3}. It was also shown in~\cite{davydov2009modular} that the condensation is fully-confining if and only if $A=H$, in which case the function $\eta$ determines $\epsilon$ through \eqref{eq:condense-3} and the condensation is determined solely by a subgroup $H<\Gamma$ and a 2-cocycle of $H$, which we recall is the same as the classification of 1+1D gapped $\Gamma$-phases, i.e. an SSB from $\Gamma$ to $H$, and an SPT of $H$.  On the other hand, if no charge is condensed and the condensation is not fully-confining, meaning that $H=\Gamma$, and $A$ is a proper subgroup of $G$, the data becomes exactly the same as what determines an igSPT: a group extension $1\to A\to \Gamma\to G:=\Gamma/A\to 1$, a 2-cocycle of $A$, and a function $\epsilon: A\times H\to U(1)$. And the conditions~\eqref{eq:condense-1}-\eqref{eq:condense-3} are exactly the same as the consistency conditions in~\eqref{eq:consistency conditions 1}-\eqref{eq:consistency conditions 3}. We conclude that there is a 1-1 correspondence between symmetric partially-confining condensations of $D(\Gamma)$ and 1+1D $\Gamma$-gSPTs. 

Moreover, the post-condensation topological order is in general a twisted quantum double given by the following:

\textit{(\cite{davydov2016lagrangian}, Thm. 2.17)
    The category of local modules over $\A(A,\eta,\epsilon)$ is equivalent to $D(G:=\Gamma/A)_\omega$ as a UMTC, with the twist $\omega$ given by:
    \begin{align}
        &\omega_3(g,h,k)=\epsilon(e_2(g,h),k^{-1})\nonumber\\
        &\times\eta(e_2(g,hk),e_2(h,k))\eta(e_2(g,hk)e_2(h,k),k^{-1}e_2(g,h)^{-1}k)\label{eq:twist}
    \end{align}
}
Although appears to be different from \eqref{eq:emergent anomaly}, it can be shown that~\eqref{eq:twist} is equivalent to the emergent anomaly of gSPT~\eqref{eq:emergent anomaly} as cocycles. We show the equivalence between the two in appendix~\ref{app:twist=anomaly}. In summary, there is a 1d-2d correspondence between gSPTs and symmetric partially-confining condensations of quantum doubles:
\begin{mdframed}
        A gSPT defined by the data $(A\lhd \Gamma,\eta,\epsilon)$ is described in SymTFT by 
        the condensable algebra $\A(A\lhd \Gamma,\eta,\epsilon)$ of $D(\Gamma)$. If the post-condensation topological order is $D(\Gamma)^{loc}_{\A}\cong D_\omega(G)$, then the emergent anomaly of the gSPT is equal to the twist $\omega\in H^3[G,U(1)]$.
\end{mdframed}
It fact it is readily seen that the symmetry of a gSPT has the structure of a $\Gamma$-graded vector space with $\Gamma$-action. The localized edge actions $\widetilde{U}_a$ span a graded vector space supported on $A$, with a $\Gamma$-action given by the conjugation $U_\gamma \widetilde{U}_aU_{\gamma}^\dagger=\epsilon(a,\gamma)\widetilde{U}_{\gamma a\gamma^{-1}}$. Therefore the symmetry algebra of a $\Gamma$-gSPT naturally defines an object of $D(\Gamma)$. The product of operators: $\widetilde{U}_a\widetilde{U}_b=\eta(a,b)\widetilde{U}_{ab}$ endows the object with an algebra structure, and the consistency conditions\eqref{eq:consistency conditions 1}-\eqref{eq:consistency conditions 3} make sure the algebra is condensable.
\section{Special cases of gSPTs and their SymTFT description\label{sec:gSPTegs}}
Having established the formal relation between anyon condensation and gSPT, let us now look at several typical classes of gSPTs and how they are described by the SymTFT as applications of this formalism. There are two families of gSPTs that have been studied in systematic manners. One is the gSPTs of SPT-pump type, another is the week gSPTs that can thought of as transition from SPT to spontaneous symmetry breaking.  We discuss how they are incorporated into our general classification in terms of the data $(\eta,\epsilon)$ as well as their SymTFT description.
\subsection{igSPT of SPT-pump type}
A class of igSPTs whose edge modes can be explained by the mechanism of SPT-pump was studied in~\cite{Wen_2023}. In terms of the data $\eta,\epsilon$, these gSPTs are solutions with $\eta=0$, and whose emergent anomaly is given by $\omega_3=\epsilon\cup e_2$. In this section we show how gSPTs of SPT-pump type are described by the SymTFT. For simplicity let us consider the case where $\Gamma$ is Abelian. The quantum double $D(\Gamma)$ that is dual to the igSPT is an Abelian gauge theory whose excitations are charges, fluxes and their bond states. In general fluxes are labelled by elements of the gauge group $\Gamma$, while charges are labelled by irreducible representations of the group, or in the abelian case the dual group $\widehat{\Gamma}:=H^1[\Gamma,U(1)]$. Therefore we denote the fluxes as $m_\gamma,\gamma\in\Gamma$ and the charges as $e_{\hat{\gamma}}, \hat{\gamma}\in\widehat{\Gamma}$. Both charges and fluxes are self-bosons, and the mutual statistics between a charge and a flux is given by the pairing between $\Gamma$ and $\widehat{\Gamma}$: $S_{e_{\hat{\gamma}},m_\gamma}=\hat{\gamma}(\gamma)\in U(1)$. From our understanding of igSPTs of SPT-pump type, we will be able to reverse engineer the corresponding anyon condensation. If the igSPT has a group extension of the form 
\begin{figure*}
    \centering
    \includegraphics[width=1.2\columnwidth]{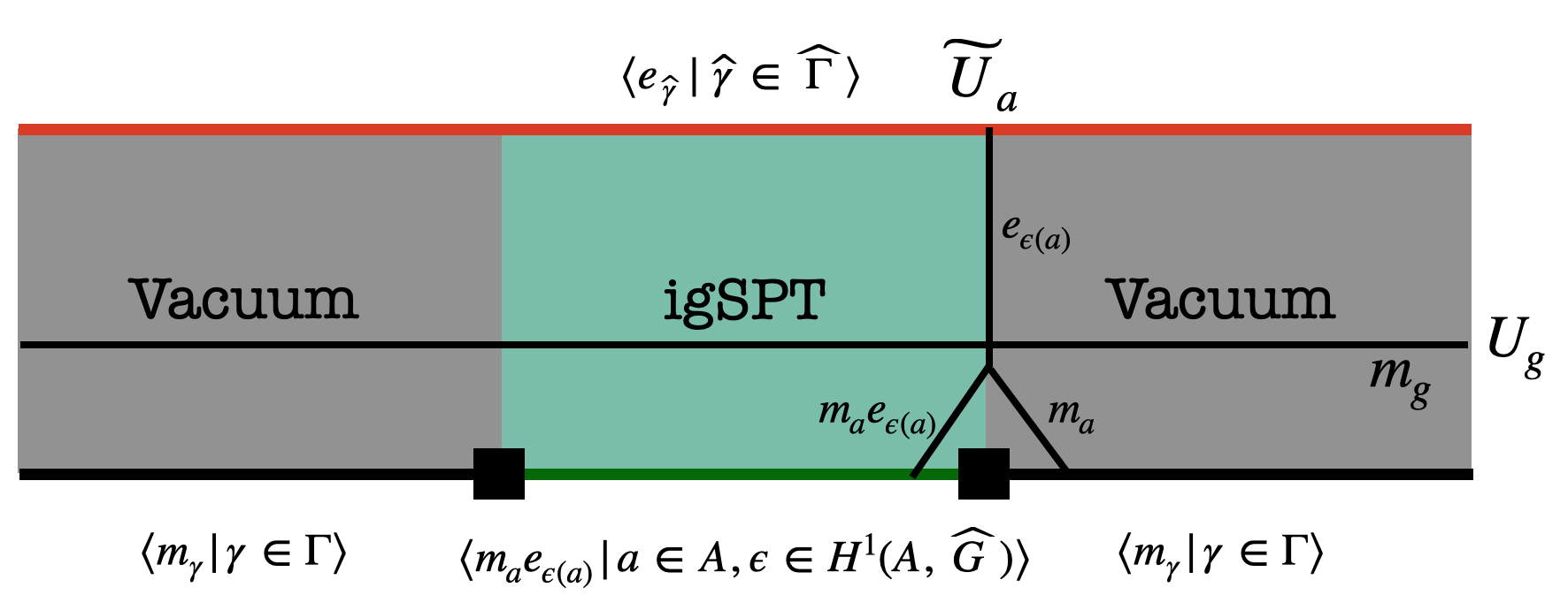}
    \caption{\textbf{Edge modes of igSPTs of SPT-pump type in the thin slab construction.} On the top reference boundary we fix the charge-condensed boundary condition. On the physical boundary we consider a condensation determined by the SPT-pump factor $\epsilon$ where fluxes of $A$ are binded with charges $\epsilon(a,-)$ and then condensed. The resulting slab has the expected SPT-pump edge modes: an $a\in A$ action is localized to the edge of the igSPT, and its charge under a $g\in G$ action is $\epsilon(a,g)$.}
    \label{fig:igSPT2}
    \end{figure*}
\begin{align}
    1\to A\to \Gamma\to G\to 1,
\end{align}
then we expect an action by $a\in A$ to pump a charge $\epsilon(a,-)$. In the $\bZ_4$-igSPT example, the $m^2$ anyon is binded with $e^2$ before condensation, which is essentially what is pumped by the $\bZ_2^A$-action. Therefore in order for $a$-action to pump a charge $\epsilon(a,-)\in \widehat{G}$~\footnote{The phase $\epsilon(a,\gamma)$ in general is defined on $A\times \Gamma$, but when $\eta=0$, the consistency condition~\eqref{eq:consistency conditions 3} states that $\epsilon|_{A\times A}=1$.  $\epsilon$ is now also linear in both variables by~\eqref{eq:consistency conditions 1}-\eqref{eq:consistency conditions 2}, therefore $\epsilon$ is now an element of $H^1[A\times G,U(1)]=H^1[A,\widehat{G}]$}, we expect the bond states $m_ae_{\epsilon(a)}$ to be condensed, where we defined $\epsilon(a):=\epsilon(a,-)$ which is an element of $\widehat{G}$, corresponding to a gauge charge.  It is readily seen that anyons $m_ae_{\epsilon(a)}$ can be condensed simultaneously: the quantum spin is $\theta(m_ae_{\epsilon(a)})=S_{m_a,e_{\epsilon(a)}}=\epsilon(a,a)=1$ since $\epsilon|_{A\times A}=1$, the mutual statistics between $m_ae_{\epsilon(a)}$ and $m_b e_{\epsilon(b)}$ is $S_{m_ae_{\epsilon(a)},m_b e_{\epsilon(b)}}=\epsilon(a,b)\epsilon(b,a)=1$. By the same diagramatic reasoning that leads to edge modes of the $\bZ_4$-igSPT, we see that the condensation $\A=\{m_ae_{\epsilon(a)},a\in A\}$ makes the $A$-actions localized to the edge while the $G$-actions are not localizable. The charges binded with $m_a$ give the expected SPT-pump structure. Therefore this condensation should be dual to the $\Gamma$-igSPT of SPT-pump type. Since we also know that these igSPTs have emergent anomalies given by $\omega_3=\epsilon\cup e_2$, we arrive at the following conclusion: 
\begin{mdframed}
    Let $A< \Gamma$ be a subgroup, $G:=\Gamma/A$, $e_2$ is the extension class associated with $1\to A\to \Gamma\to G\to 1$. For any $\epsilon\in H^1[A,\widehat{G}]$, condensing $\A=\{m_ae_{\epsilon(a)},a\in A\}$ in the $D(\Gamma)$ quantum double results in the topological order $D_\omega(G=\Gamma/A)$, with $\omega=\epsilon\cup e_2$.
\end{mdframed}
We see that via the SymTFT/igSPT duality and by invoking our knowledge about igSPTs in 1+1D, we have inferred information regarding condensations in 2+1D quantum doubles. See Fig.~\ref{fig:igSPT2} for an illustration of the thin slab construction of igSPTs of SPT-pump type. 

Similar to the $\bZ_4$-igSPT, igSPTs of SPT-pump types are all stable due to their emergent anomalies. From the anyon condensation perspective, this means it is impossible to further condense anyons to arrive at a fully-confining condensation that does not condense charges. This can be seen from that fact that the post-condensation topological order is a twisted quantum double, and it is well known that in a twisted quantum double there is no fully-confining condensation that does not condense charges~\cite{zhang2023anomalies}. Therefore in order to arrive at a gapped symmetric phase one has to either condense charges, corresponding to breaking the symmetry, or ``uncondense'' some anyons in $\A$, which corresponds to closing the gap to $A$ degrees of freedom. This anyon condensation picture matches with our expectation from the 1+1D analysis: as the gap to $A$-dof is perturbatively stable, closing the gap to $A$ degrees of freedom will require a phase transition out of the igSPT phase. On other hand the anomaly of the gapless sector forbids opening a gap directly without breaking the symmetry. 

\subsection{Weak gSPTs}
There is also a family of gSPTs named weak gSPTs that are transitions between SPT phases and symmetry breaking phases, these have been studied in~\cite{Verresen_2021,li2023decorated}. Here we discuss some generalities of such gSPTs and show how they fall into our classification in terms of the pair $(\eta,\epsilon)$ as well as their SymTFT description. The common mechanism for such gSPTs is that the full symmetry has the form $G^A\times G^B$, and on one side of the transition is an SPT of decorated domain wall(DDW) type, where domain walls of $G^A$ are decorated with charges of $G^B$ and vice versa. On the other side of the transition the symmetry is broken down to $G^A$, and system is a $G^B$-paramagnet. The transition between the two is thus in the same universality class as a usual $G^B$-SSB transition, which is further enriched by $G^A\times G^B$ to form a gSPT.  The $G^A$ symmetry is intact in both the SPT and the $G^B$-SSB phases, therefore acts as a gapped symmetry at the transition point. Thus $G^A$-action still localizes to edge actions at the transition, and the restriction to one edge carries charges of $G^B$ from the DDW structure of the SPT phase. This leads to edge modes and makes the transition a nontrivial gSPT. On the other hand the $G^B$-action, which acts on gapless dof, no longer fractionalizes and the edge modes of the gSPT are reduced from that of the nearby $G^A\times G^B$-SPT. As the group extension is trivial, there is no emergent anomaly. The gSPT can be gapped out directly while preserving symmetry-it is adjacent to a gapped SPT by construction. Thus such gSPTs are not stable, and the edge modes of them are nothing but residue of SPT edge modes, hence the name weak gSPTs. Nevertheless, the symmetry enriching pattern and nontrivial edge modes make it distinct from a usual $G^B$-SSB transition with a trivial $G^A$-action. In particular, the charge carried by edge $G^A$-action provides a discrete invariant that can not be altered without changing the bulk universality class. To go to a trivial $G^B$-SSB transition the system has to either open up a gap or go through a multi-critical point with a different bulk universality class. 

Examples of weak gSPTs include the Haldane chain to SSB transition, which has an Ising CFT bulk and a 2-fold GSD on an opne chain, and similarly $Z_3\times Z_3$-SPT to SSB transitions, which have bulk three-state Potts CFT with central charge c = 4/5 and 3-fold GSD on an open chain~\cite{Verresen_2021}. Numerical results on the phase diagrams of these models have been obtained in~\cite{Verresen_2021}, in either case it was confirmed that the weak gSPT is separated from a trivial one by a multi-critical point. We will later provide a rather simple derivation of the phase diagram using SymTFT.

Let us now see how the weak gSPTs fall into our classification in terms of $\eta,\epsilon$. A weak gSPT has a symmetry group $G^A\times G^B$, and on one side of the transition is in an SPT of DDW type. In the cohomology classification this means the SPT is described by a cocycle $\beta_2\in H^1[G^A,H^1[G^B,U(1)]]$, which is a subgroup of the full cohomology group $H^2[G^A\times G^B,U(1)]$ by the Künneth formula. This means $\beta_2=\beta_2(g^A,h^B)$ is linear in both factors and describes the charges carried by edge $G^A$ and $G^B$ actions via 
\begin{align}
    &U_{g^B}\widetilde{U}_{g^A}U_{g^B}^\dagger=\beta_2(g^A,g^B)\widetilde{U}_{g^A}\\
    &U_{g^A}\widetilde{U}_{g^B}U_{g^A}^\dagger=\beta_2(g^A,g^B)^{-1}\widetilde{U}_{g^B}
\end{align}
At the transition point the subgroup $G^B$ becomes gapless and the second equation above is lost. However we still have the first equation from which we see that it corresponds to a gSPT with $\eta_2=1,\epsilon(a=g^A,\gamma=(h^A,h^B))=\beta_2(g^A,h^B)$.~\footnote{We note that a weak gSPT can be defined more broadly as any SPT to (partial)SSB transition. Let $\beta_2\in H^2[\Gamma,U(1)]$ represent a $\Gamma$-SPT. If one drives the SPT to the $\Gamma$ broken down to $A\unlhd \Gamma$ transition, then the transition is a gSPT with extension $1\to A\xrightarrow{i} \Gamma\to G\to 1$ and  $\eta=i^*(\beta_2), \epsilon(a,\gamma)=\frac{\beta_2(\gamma,a)}{\beta_2(\gamma a\gamma^{-1},\gamma)}$.}

We next discuss the SymTFT description of weak gSPTs. We take $\Gamma=\bZ_2\times \bZ_2$ as an example, and generalization to generic weak SPTs is straightforward. Since the full symmetry is $\bZ_2\times \bZ_2$, we search for symmetric partially-confining condensations that do not condense charges in $D(\bZ_2\times \bZ_2)$, i.e. two copies of toric code. One choice is the $\A=\{e_1m_2\}$ condensation. The post-condensation topological order is generated by $e_2m_1$ and $e_1$, both of which are bosons and these is a mutual $\pi$ statistics between them. Thus the post-condensation topological order is equivalent to the toric code, $D(\bZ_2^2)/\A\cong D(\bZ_2)$, and the condensation describes a 1+1D gSPT with a trivial group extension $1\to \bZ^A_2\to \bZ_2^A\times \bZ_2^B\to \bZ_2^B\to 1$. Using the thin slab construction, it is readily seen that $\bZ_2^A$-action is localized to the edge and the edge action is charged under the $\bZ_2^B$-action. In terms of the pair $(\eta,\epsilon)$, this means we have $\eta=1, \epsilon(a,\vec{b})=(-1)^{ab^1}$. Since the extension is trivial, there is no emergent anomaly. Nevertheless, the nontrivial $\epsilon$ leads to a 2-fold GSD on an open chain. Starting with the condensation $e_1m_2$, further condensing $e_2m_1$ is allowed and will result in the condensation $\langle e_1m_2,e_2m_1\rangle$, which is dual to the cluster chain. On the other hand further condensing $e_1$ or $m_2$ will result in the condensation $\langle e_1,m_2\rangle$, which is dual to a $\bZ_2^A$-SSB/$\bZ_2^B$-paramagnet phase. Therefore the condensation $\langle e_1m_2\rangle$ describes the transition from the cluster chain to the $\bZ_2^A$-SSB/$\bZ_2^B$-paramagnet phase. The trivial gSPT with the same bulk universality class is described by the condensation $\langle m_2\rangle$, which corresponds to the $\bZ_2^A\times\bZ_2^B$-paramagnet to $\bZ_2^A$-SSB/$\bZ_2^B$-paramagnet transition.  From the SymTF it is clear that the two gSPTs are not connected if symmetry is preserved. Both gSPTs are in the Ising-CFT universality class, but in order to go from the $e_1m_2$ condensation to the $m_2$ condensation one has to either condense charges, corresonding to breaking the symmetry, or uncondense $e_1m_2$, which will lead to the $\{1\}$ condensation that is dual to a $\bZ_2^A\times\bZ_2^B$-multi-critical point. The same phase diagram has been obtained in~\cite{Verresen_2021} via DMRG.

\subsection{Stacking gSPT with SPT\label{sec:SPT_stack}}
One can always stack a gSPT with an SPT to obtain a new gSPT. We discuss some generalities of this kind of construction in this section.

\subsubsection{An example}
Consider the topological order $D(\bZ_4^A\times \bZ_4^B)$, which is two copies of $\bZ_4$-Toric codes. Denote the charges and fluxes in each copy as $e_{1,2}$ and $m_{1,2}$. Consider the condensation $\A=\<m_1^2e_1^2e_2^2,m_2e_1\>$. The post-condensation topological order consists of the deconfined anyons $m_1e_1e_2^3, m_1e_1^3e_2^3$ and $m_2^2$, which are semion, anti-semion and boson respectively. We conclude that the condensation leaves a twisted quantum double $D_\omega(\bZ_2)$. Let us consider the slab construction with the condensation $\A$ on the physical boundary. Following the same arguement as in Fig.~\ref{fig:SPT_edge1}, one can see that the $\bZ_2^a$ subgroup of $\bZ_4^a$ and the full $\bZ_4^b$ subgroup can be localized to the edge. This corresponds to a gapped symmetry $A=\bZ_2^a\times \bZ_4^b$, giving a group extension is $1\to \bZ_2^a\times \bZ_4^b\to \bZ_4^a\times \bZ_4^b\to \bZ_2\to 1$. However now the edge $\bZ_2$ action and edge $\bZ_4$ action of $A$ anti-commute. This property is similar to the symmetry action of a non-trivial $\bZ_2\times \bZ_4$-SPT. We conclude that this condensation corresponds to a gSPT with a non-trivial $\eta\in H^2[\bZ_2\times\bZ_4,U(1)]\cong \bZ_2$. Also, the action by both $\bZ_2^a$ and $\bZ_4^b$ of $A$ on the edge anti-commute with the $G$-action, i.e. a horizontal $m_1$-string. Therefore this condensation/gSPT also has a non-trivial $\epsilon$ factor. 

The condensation and the thin slab construction provide a complete description of the dual gSPT: the group extension, the edge modes and the emergent anomaly. As a consistency check, one can verify that $\eta(\vec{a},\vec{b})=(-1)^{a^1b^2},~\epsilon(\vec{a},\vec{\gamma})=(-1)^{a^1\gamma^1+a^1\gamma^2}i^{a^2\gamma^1}$ is a solution to the consistency conditions on $(\eta,\epsilon)$, and by inspection this solution gives a gSPT with the same structure as that is given by the condensation $\A=\<m_1^2e_1^2e_2^2,m_2e_1\>$.

Notably this igSPT can be viewed as obtained by stacking the $\bZ_4$-igSPT with a $\bZ_4\times\bZ_4$-SPT. In other words the $A$-SPT given by $\eta$ can be cancelled by stacking it with a $\Gamma$-SPT. In this case $\Gamma=\bZ_4\times \bZ_4$-SPTs have a $\bZ_4$ classification. Take the generating phase, the pull-back of this $\Gamma$-SPT to the subgroup $A=\bZ_2\times\bZ_4$ is exactly the non-trivial $\bZ_2\times \bZ_4$-SPT, therefore cancels the $A$-SPT of this igSPT upon stacking. The SPT+igSPT will still have the symmetry extension $1\to \bZ_2^a\times \bZ_4^b\to \bZ_4^a\times \bZ_4^b\to \bZ_2\to 1$, but now the second factor $\bZ_4^b$ of $\Gamma$ plays no role and the system is equivalent to the $\bZ_4^a$-igSPT with an extra trivial $\bZ_4^b$-action. 

\subsubsection{General consideration}
Let us now consider the effect on the data $(\eta,\epsilon)$ of stacking with an SPT specified by $\omega_2\in H^2[\Gamma,U(1)]$. For simplicity we assume all groups are abelian here, generalization to non-abelian groups is straightforward. The stacked system has symmetry $\Gamma^{SPT}\times \Gamma^{gSPT}$, but we are only interested in the diagonal symmetry $\Gamma$ and can assume the symmetry is broken down to this diagonal subgroup by perturbations.  Now the symmetry extension is unchanged by the stacking, because still only the $A$-subgroup acts trivially on the gapless sector, and the quotient $G=\Gamma/A$ acts nontrivially on the gapless sector coming from the gSPT. The $A$-action is now modified by the $\Gamma$-SPT, namely $\eta$ is now modified by $j^*(\omega_2)$, where $j$ is the inclusion $A\xrightarrow{j} \Gamma$ and $j^*$ is the pull-back of $\omega_2$ to $A$. Recall the factor $\epsilon$ is the measurement of the charge of the edge $A$-actions under $\Gamma$. An SPT also defines such a factor: the edge of an $a\in A$ action carries charge $\frac{\omega_2(\gamma,a)}{\omega_2(a,\gamma)}$ under $\gamma$. Therefore the factor $\epsilon$ is modified by $i_a\omega(\gamma)$, where we used the slant product $i_a\omega(\gamma):=\frac{\omega_2(\gamma,a)}{\omega_2(a,\gamma)}$. Notice the set of data $(j^*(\omega_2),i_a\omega_2(\gamma))$ satisfies the consistency conditions~\eqref{eq:consistency conditions 1}-\eqref{eq:consistency conditions 3}. The product of two solutions is again a solution, therefore the data $(\eta j^*(\omega_2),\epsilon(a,\gamma) i_a\omega_2(\gamma))$ is a valid solution and corresponds to the gSPT obtained by stacking the $(\eta,\epsilon)$-gSPT with the $\omega_2$-SPT. It can also be shown by direct computation that two gSPTs differ by stacking with an SPT have the same emergent anomaly.

In the $\bZ_4^a\times \bZ_4^b$-gSPT example studied earlier, one can verify that choosing the $\omega_2$ corresponding to the generating element of $H^2[\bZ_4\times \bZ_4,U(1)]$ will lead to a pull-back $j^*(\omega_2)$ that is equal to the generating element of $H^2[\bZ_2\times \bZ_4,U(1)]$, therefore trivializing the $\eta$ factor of the $\bZ_4^a\times \bZ_4^b$-gSPT upon stacking, leaving an igSPT of SPT-pump type.

\section{Discussion and outlook\label{sec:discussion}}
In this work  established a one-to-one correspondence between 1+1D bosonic gapless SPTs (gSPTs) and symmetric, partially-confining anyon condensations in 2+1D quantum doubles. 
These results suggest a number of directions for future exploration. 
Fermionic igSPTs have been studied in~\cite{Thorngren_2021}, including an example that is similar to the bosonic $\bZ_4$-igSPT but with fermion parity playing the role of the gapped symmetry group. Can one construct a SymTFT dual to such fermionic igSPTs? This would first require understanding how to incorporate fermion parity, which differs in important ways from ordinary zero-form symmetries, into the SymTFT description.

Moreover, it is natural to consider gSPTs protected by generalized symmetries.
In 1+1D non-invertible symmetries are described by fusion categories. Anomalies of non-invertible symmetries have been studied in~\cite{kaidi2023symmetry,zhang2023anomalies,thorngren2019fusion,bhardwaj2023gapped}. The SymTFT for 1+1D non-invertible symmetry is also well understood, if the symmetry forms a fusion category $\EA$, then the SymTFT is given by the center $\Z[\EA]$, which can also be viewed as the Levin-Wen string-net model~\cite{Levin_2005} with input $\EA$. Following the SymTFT dictionary, a condensation of $\Z[\EA]$ that i) does not confine the topological order completely ii) does not condense generalized charges~\cite{zhang2023anomalies} would be dual to a gSPT protected by $\EA$. The post-condensation topological order is in general another Levin-Wen model $\Z[\EB]$. If $\EA$ is itself gappable while $\EB$ is not, then we can say that the gSPT has an emergent anomaly. It would be interesting to extend the algebraic analysis of symmetries we performed for gSPTs in this work to non-invertible symmetries and compare with the known classification of anyon condensations of Levin-Wen models~\cite{Kitaev_2012}. 

In higher dimensions it is also natural to consider gSPTs protected by higher-form symmetries or mixture of ordinary- and higher-form- symmetries. SymTFT for higher form symmetries have also been studied~\cite{Kong_2020,Chatterjee_2023_1} and can be viewed as (twisted) higher-form gauge theories.  In fact, it is known that gauging a subgroup $A\lhd \Gamma$ of a 0-form symmetry results in a symmetry $\Gamma/A\times \widehat{A}^{D-2}$, where $\widehat{A}^{D-2}$ is a $D-2$-form symmetry generated by Wilson loops of the $A$-gauge field. There is a mixed anomaly between $\Gamma/A$ and $\widehat{A}^{D-2}$~\cite{Tachikawa_2020}. One can replace the hard gauging where the Hilbert space only consists of gauge invariant states, by the so called soft gauging~\cite{Borla_2021,thorngren2023higgs,verresen2022higgs} where Gauss's law is imposed only energetically. In this way, the mixed anomaly becomes emergent, with a symmetry extension structure $1\to A\to \Gamma\times\widehat{A}^{D-2}\to \Gamma/A\times \widehat{A}^{D-2}\to 1$. Therefore soft gauging a subgroup gives a gSPT protected by a mixture of zero and higher-form symmetry in spacetime dimension greater than two. A similar construction was discussed in~\cite{su2023boundary}. It would be interesting to see if a correspondence between gSPTs and SymTFT holds in the presence of higher form symmetries, and to extend the classification to gSPTs protected by such symmetries. 

\vspace{12pt}\noindent {\it Acknowledgements -- }
We thank Weicheng Ye, Joseph Sullivan, Fiona Burnell, and Julio Parra-Martinez for insightful discussions.
This work is supported by the Natural Sciences and Engineering Research Council of Canada (NSERC), and by an Alfred P. Sloan Foundation Fellowship (A.C.P.). We thank the Aspen Center for Physics where part of this work was completed

 \vspace{12pt}\noindent{\it Note -- }
During the completion of this manuscript, we became aware of Ref.~\cite{huang2023topological}, which examines SymTFT duals gapless states and critical points and has some overlap with our results.

\bibliography{hologSPT.bib}
\clearpage
\appendix
\section{Consistency conditions on the data $\eta,\epsilon$\label{app: consistency conditions}}
First of all we can compute $U_\gamma \widetilde{U}_a\widetilde{U}_b U_\gamma^{-1}$ in two ways,
\begin{align}
    U_\gamma \widetilde{U}_a\widetilde{U}_b U_\gamma^{-1}=\eta(a,b)U_\gamma\widetilde{U}_{ab}U_\gamma^{-1}=\eta(a,b)\epsilon(ab,\gamma)\widetilde{U}_{\gamma ab\gamma^{-1}},
\end{align}
we also have
\begin{align}
    U_\gamma \widetilde{U}_a\widetilde{U}_b U_\gamma^{-1}&=U_\gamma \widetilde{U}_{a}U_\gamma^{-1}U_\gamma\widetilde{U}_bU_\gamma^{-1}\\
    &=\epsilon(a,\gamma)\epsilon(b,\gamma)\widetilde{U}_{\gamma a\gamma^{-1}}\widetilde{U}_{\gamma b\gamma^{-1}}\nonumber\\
    &=\epsilon(a,\gamma)\epsilon(b,\gamma)\eta(\gamma a\gamma^{-1},\gamma b\gamma^{-1})\widetilde{U}_{ab},
\end{align}
we conclude 
\begin{align}
   \eta(a,b)\epsilon(ab,\gamma)=\eta(\gamma a\gamma^{-1},\gamma b \gamma^{-1})\epsilon(a,\gamma)\epsilon(b,\gamma)
\end{align}
Secondly we can compute 
\begin{align}
    &U_{\gamma_1}U_{\gamma_2}\widetilde{U}_aU_{\gamma_2}^{-1}U_{\gamma_2}^{-1}\nonumber\\
    &=\epsilon(a,\gamma_2)U_{\gamma_1}\widetilde{U}_{\gamma_2a\gamma_2^{-1}}U_{\gamma_1}^{-1}\\
    &=\epsilon(a,\gamma_2)\epsilon(\gamma_2 a\gamma_2^{-1},\gamma_1)\widetilde{U}_{\gamma_1\gamma_2 a\gamma_2^{-1}\gamma_1^{-1}}.
\end{align}
On the other hand 
\begin{align}
    &U_{\gamma_1}U_{\gamma_2}\widetilde{U}_aU_{\gamma_2}^{-1}U_{\gamma_2}^{-1}\nonumber\\
    &=U_{\gamma_1\gamma_2}\widetilde{U}_aU_{\gamma_1\gamma_2}=\epsilon(a,\gamma_1\gamma_2) \widetilde{U}_{\gamma_1\gamma_2 a\gamma_2^{-1}\gamma_1^{-1}}.
\end{align}
We conclude that $\epsilon$ satisfies:
\begin{align}
    \epsilon(a,\gamma_1\gamma_2)=\epsilon(\gamma_2 a\gamma_2^{-1},\gamma_1)\epsilon(a,\gamma_2).
\end{align}
Lastly, there are two ways of computing $U_\gamma \widetilde{U}_a U_\gamma^{-1}$ when $\gamma=b$ is an element of $A$:
\begin{align}
    U_b\widetilde{U}_{a}U_b^{-1}&=\widetilde{U}_b\widetilde{U}_a\widetilde{U}_b^{-1}=\eta(b,a)\widetilde{U}_{ba}\widetilde{U}_{b}^{-1}\nonumber\\
    &=\eta(b,a)\eta(ba,b^{-1})\widetilde{U}_{bab^{-1}}\nonumber\\
    &=\frac{\eta(b,a)}{\eta(bab^{-1},b)}\widetilde{U}_{bab^{-1}},
\end{align}
where we used $\eta(ba,b^{-1})=1/\eta(bab^{-1},a)$ which comes from cocycle condition on $\eta$ (we assume cocycles are normalized so that $\eta(1,a)=\eta(a,a^{-1})=1$). Since we also have
\begin{align}
    U_b\widetilde{U}_{a}U_b^{-1}=\epsilon(a,b)\widetilde{U}_{bab^{-1}},
\end{align}
we conclude that $\epsilon(a,b)=\frac{\eta(b,a)}{\eta(bab^{-1},b)},~\forall a,b\in A$. 
\section{Connection to the LHS spectral sequence\label{app:ss}}
Here we show that the consistency conditions~\eqref{eq:consistency conditions 1}-\eqref{eq:consistency conditions 3} as well as the emergent anomaly~\eqref{eq:emergent anomaly} have the structure of the LHS spectral sequence. We focus on computation aspects of the LHS sequence and work with cochain-level expressions. We also assume the group $A$ is abelian in this appendix.

Consider a 3-cocycle of $G$: $\omega_3\in Z^3[G,U(1)]$ that is trivialized via a group extension 
\begin{align}
    1\rightarrow A\xrightarrow[]{j} \Gamma\xrightarrow[]{p} G\rightarrow 1.\label{eq: groupext2}
\end{align}
This means that there is some 2-cochain of $\Gamma$: $\alpha\in \C^2[\Gamma,U(1)]$, such that $d\alpha((a_1,g_1),(a_2,g_2),(a_3,g_3))=\omega(g_1,g_2,g_3)$. Here we have written elements of $\Gamma$ as pairs $(a,g), a\in A,G\in G$. We also denote the action of $G$ on $A$ by $^ga:=gag^{-1}$.The group extension~\eqref{eq: groupext2} specifies an extension class $e_2\in \Z^2[G,A]$. By modifying with coboundaries one can always decompose the 2-cochain $\alpha$ into the following form~\cite{wang2021domain}:
\begin{align}
    &\alpha((a,g),(b,h))\nonumber\\
    &=F^{0,2}(^{\bar{g}\bar{h}}a,^{\bar{h}}b)F^{0,2}(^{\bar{g}\bar{h}}e_2(g,h),^{\overline{gh}}a^{\bar{h}}b)F^{1,1}(^{\bar{g}}a,h)F^{2,0}(g,h),\label{eq:alphadec}
\end{align}
where $F^{0,2}\in \C^2[A,U(1)]$, $F^{1,1}\in \C^1[A,\C^1[G,U(1)]]$, $F^{2,0}\in \C^2[G,U(1)]$.
Requiring that $d\alpha((a,g),(b,h),(c,k))$ only depends on $g,h,k$ results in the following conditions:
\begin{align}
    &\delta_0F^{0,2}=1,\\
    &\delta_1F^{0,2}\delta_0F^{1,1}=1,\\
    &\delta_2F^{0,2}\delta_1F^{1,1}\delta_0F^{0,2}=1.
\end{align}
Here the differentials $\delta_i$ are defined as follows. $\delta_0$ is the cochain differential with respect to $A$: $\delta_0=d_A$, $\delta_1$ is the cochain differential with respect to $G$: $\delta_1=d_G$, and 
\begin{align}
    \delta_2F_{0,2}(a,g,h):=\frac{F^{0,2}(a,e_2(g,h))}{F^{0,2}(e_2(g,h),a)}.
\end{align}
Once these conditions are met, the trivializing relation $\omega(g,h,k)=d\alpha((a,g),(b,h),(c,k))$ becomes 
\begin{align}
    \omega(g,h,k)=\delta_3 F^{0,2}\delta_2 F^{1,1}\delta_1 F^{2,0}
\end{align}
where $\delta_3, \delta_2$ are defined by 
\begin{align}
    &\delta_3F^{0,2}(g,h,k):=\frac{F^{0,2}(e_2(g,hk),^ge_2(h,k))}{F^{0,2}(e_2(gh,k),e_2(g,h))}\\
&\delta_2 F^{1,1}(g,h,k):=F^{1,1}(e_2(g,h),k)
\end{align}
Since $\delta_1=d_G$, we see that on the cocycle level we can neglect $F^{2,0}$ and $\omega\cong \delta_3 F^{0,2}\delta_2F^{1,1}$ as $G$-cocycles. Now we can identify $F^{0,2}(a,b)=\eta(b,a)$ and $F^{1,2}=\epsilon(a,g)$. The condition that $\eta$ is a 2-cocycle of $A$ is the same as $\delta_0 F^{0,2}=1$. The consistency condition~\eqref{eq:consistency conditions 1} is
\begin{align}
    \frac{\epsilon(a,g)\epsilon(b,g)}{\epsilon(ab,g)}=\frac{\eta(a,b)}{\eta(^ga,^gb)}
\end{align}
which is the same as $d_A\epsilon d_G\eta=1$, or $\delta_0F^{0,2}\delta_1F^{1,1}=1$. Lastly using the condition~\eqref{eq:consistency conditions 2},
\begin{align}
    \epsilon(^ha,g)\epsilon(a,h)=\epsilon(a,ghe_2(g,h))=\epsilon(a,gh)\frac{\eta(a,e_2(g,h))}{\eta(e_2(g,h),a)}
\end{align}
which is the same as $\delta_1F^{1,1}\delta_2F^{0,2}=1$. The emergent anomaly~\eqref{eq:emergent anomaly} can be readily seen to be the same as $\delta_3 F^{0,2}\delta_2 F^{1,1}$. Since now $d_\Gamma \alpha=\delta_3 F^{0,2}\delta_2 F^{1,1}$, this means the anomaly ~\eqref{eq:emergent anomaly} is the differential of the 2-cochain $\alpha$, therefore is trivialized by the group extension.
\section{The bulk-edge correspondence for 1+1D gSPTs\label{app: gSPTbulkedge}}
A 1+1D gSPT is defined by a group extension~\eqref{eq: groupext2} and a pair $(\eta,\epsilon)$. The emergent anomaly is determined by $(\eta,\epsilon)$ via~\eqref{eq:emergent anomaly}. This can be viewed as an edge to bulk map: the edge modes determines the bulk emergent anomaly. To show the converse the question is equivalent to finding $\eta,\epsilon$ for a given emergent anomaly. To start with, one finds a trivializing cochain: $d\alpha=\omega, \alpha\in \C^2[\Gamma,U(1)]$. From the decomposition \eqref{eq:alphadec} we see that $\alpha((a,1),(b,1))=F^{0,2}(a,b)$. Formally this means the pull back of $\alpha$ by $j$ is a 2-cocycle of $A$: $dj^*(\alpha)=j^*d\alpha=j^*p^*\omega=(p\circ j)^*\omega=1$, where we used $(p\circ j)^*=1$ because the sequence \eqref{eq: groupext2} is exact. Similarly, other terms in the decomposition $F^{1,1},F^{0,2}$ can be extracted from $\alpha$ by the Lyndon's algorithm~\cite{Lyndon}. For example 
\begin{align}
   \epsilon(a,g)=F^{1,1}(a,g)=i_a \alpha(g):=\frac{\alpha((a,1),(1,g))}{\alpha((1,g),(a,1))}.
\end{align}
Therefore $\epsilon,\eta$ can be reconstructed once the trivializing cochain $\alpha$ is found. However, the choice of the trivializing cochain is not unique: if $d\alpha=d\alpha'=\omega$, then $d(\alpha\alpha'^{-1})=1$. Thus any two $\Gamma$-cochains $\alpha,\alpha'$ that differ by a $\Gamma$-cocycle give the same emergent anomaly $\omega$. Therefore the data $\eta,\epsilon$ can only be determined from $\omega$ up to a 2-cocycle of $\Gamma$. Consider another trivializing cochain $\gamma'=\gamma\delta$, where $\delta\in \Z^2[\Gamma,U(1)]$. Then the data $\eta',\epsilon'$ associated with $\gamma'$ is 
\begin{align}
    &\eta'=j^*(\alpha')=j^*\alpha\cdot j^*\delta=\eta\cdot j^*\delta,\label{eq:newdata1}\\
    &\epsilon'(a,g)=i_a\alpha'(g)=i_a\alpha(g)\cdot i_a\delta(g)=\epsilon(a,g)i_a\delta(g)\label{eq:newdata2}
\end{align}
Recall from section~\ref{sec:SPT_stack} that stacking a gSPT defined by $(\eta,\epsilon)$ with an SPT defined by $\delta\in \Z^2[\Gamma,U(1)]$ results in a new gSPT with $\eta'=\eta\cdot j^*\delta, \epsilon'(a,g)=\epsilon(a,g)\cdot i_a\delta(g)$. Comparing with ~\eqref{eq:newdata1}-\eqref{eq:newdata2}, we see that the new data $\eta',\epsilon'$ associated with the new trivializing cochain $\alpha'$ differs from the original data $\eta,\epsilon$ associated with $\alpha$ by stacking with a $\Gamma$-SPT given by $\delta$. Therefore we conclude that for a given emergent anomaly $\omega$, the gSPT is only determined by $\omega$ up to stacking with $\Gamma$-SPTs. The fact that stacking with SPTs does not change the emergent anomaly may not very surprising, given that SPTs do not have bulk anomalies, what is surprising is that all gSPTs with the same emergent anomaly are obtained in this way. Therefore the gSPT bulk-edge correspondence can stated as follows: The edge modes of a gSPT determine the bulk emergent anomaly, and the bulk emergent anomaly determines the edge modes of the gSPT up to stacking with SPTs.

Now with the help of the bulk-edge correspondence we can prove that the two definitions of igSPT are equivalent: 1. A gSPT is intrinsic if the edge modes can not be trivialized by stacking with an SPT. 2. A gSPT is intrinsic if it has a nontrivial emergent anomaly. 1$\Rightarrow$ 2: if the emergent anomaly is trivial: $\omega=1$, then $\eta_0=\epsilon_0=1$ is clearly a solution to the consistency conditions. By the bulk-edge correspondence any other solution with the same $\omega=1$ must differ from $\eta_0=\epsilon_0=1$ by stacking with a $\Gamma$-SPT. Therefore if the gSPT has non-trivial edge modes, $\eta\neq 1$ and/or $ \epsilon\neq 1$, they can be eliminated by stacking with a $\Gamma$-SPT. $2\Rightarrow 1$: If the edge modes of a gSPT can be trivialized by stacking with a $\Gamma$-SPT, then it means the data $\eta,\epsilon$ differs from the tivial one $\eta_0=1,\epsilon_0=1$ by stacking with a $\Gamma$-SPT. Since stacking with $\Gamma$-SPTs does not change the emergent anomaly, the data $\eta,\epsilon$ gives the same emergent anomaly as that is given by the trivial one, $\omega=1$. $\Box$
\section{Equivalence between the post-condensation twist and  gSPT emergent anomaly\label{app:twist=anomaly}}
In the computation of emergent anomaly, we defined $U_gU_h=\Omega(g,h)U_{gh}$. Alternatively one can define $U_gU_h=U_{gh}\Omega'(g,h)$. Similarly $\Omega'(g,h)=\Omega'^{L}(g,h)\otimes \Omega'^R(g,h)$ The associativity condition on $\Omega'$ is
\begin{align}
    \Omega'(gh,k)U^{-1}_k \Omega'(g,h)U_k=\Omega'(g,hk)\Omega'(h,k)
\end{align} 
Restrict to $\Omega'^L$, the associativity condition is then only satisfied up to a phase,
\begin{align}
    \omega'(g,h,k)\Omega'^L(gh,k)U^{-1}_k \Omega'^L(g,h)U_k=\Omega'^L(g,hk)\Omega'^L(h,k)\label{eq:anomaly5}
\end{align}
The $\Omega'(g,h)$ is related to $\Omega(g,h)$ via $\Omega'(g,h)=U^\dagger_{gh}\Omega(g,h)U_{gh}$. Using this relation one can show that $\omega'=\omega$:
\begin{align}
    &\Omega'^L(gh,k)U^{-1}_k \Omega'^L(g,h)U_k\nonumber\\
    &=U^{-1}_{ghk}\Omega^L(gh,k)U_{ghk}U_k^{-1}U^{-1}_{gh}\Omega^L(g,h)U_ghU_k\nonumber\\
    &=U^{-1}_{ghk}\Omega^L(gh,k)\Omega^L(g,h)U_{{ghk}}
\end{align}
and 
\begin{align}
    &\Omega'^L(g,hk)\Omega'^L(h,k)=U^{-1}_{ghk}\Omega^L(g,hk)U_{ghk}U^{-1}_{hk}\Omega^L(h,k)U_{hk}\nonumber\\
    &=U^{-1}_{ghk}(\Omega^L(g,hk)U_g\Omega^L(h,k)U_g^{-1})U_{ghk}.
\end{align}
Comparing with \eqref{eq: anomaly2}, one sees that $\omega'=\omega$. Next notice that $\Omega'^L(g,h)=U^\dagger_{gh}\Omega(g,h)^LU_{gh}=U^\dagger_{gh}\widetilde{U}_{e_2(g,h)}U_{gh}=\epsilon(e_2(g,h),gh)\widetilde{U}_{e_2(g,h)}$. Therefore replacing $\Omega'L(g,h)$ by $\widetilde{U}_{e_2(g,h)}$ in~\eqref{eq:anomaly5} will only affect the cocycle $\omega$ by a coboundary (since every $\Omega'^L(g,h)$ factor only gets modified by a phase factor $\epsilon(e_2(g,h),gh)$). Therefore we can take $\Omega'^L(g,h)=\widetilde{U}_{e_2(g,h)}$ in ~\eqref{eq:anomaly5}. Write \eqref{eq:anomaly5} as 
\begin{align}
    \omega(g,h,k)\widetilde{U}_{e_2(gh,k)}=\widetilde{U}_{e_2(g,hk)}\widetilde{U}_{e_2(h,k)}U^{-1}_k (\widetilde{U}_{e_2(g,h)})^{-1}U_k.
\end{align}
Use $\widetilde{U}_a\widetilde{U}_b=\eta(a,b)\widetilde{U}_{ab}$, together with $U_g \widetilde{U}_aU_g^{-1}=\epsilon(a,g)\widetilde{U}_{gag^{-1}}$, we arrive at the expression 
\begin{align}
    &\omega(g,h,k)=\epsilon(e_2(g,h),k^{-1})\nonumber\\
    &\times\eta(e_2(g,hk),e_2(h,k))\eta(e_2(g,hk)e_2(h,k),k^{-1}e_2(g,h)^{-1}k)
\end{align}
which is the twist in~\eqref{eq:twist}. Therefore the twist~\eqref{eq:twist} and the anomaly~\eqref{eq:emergent anomaly} are equivalent as cocycles of $G$.
\end{document}